\documentclass[12pt,ams]{article}%
\usepackage[nohead]{geometry}
\usepackage{geometry}
\usepackage{dsfont}
\usepackage{amsmath}
\usepackage{amsfonts}
\usepackage{amssymb}
\usepackage{graphicx}%
\usepackage{subfig}
\usepackage{float}
\usepackage{enumitem}

\usepackage{epsfig}
\usepackage{feyn}

\usepackage{hyperref}
\begin{document}

\newcommand{\dslash}{\partial\!\!\!/}
\newcommand{\aslash}{A\!\!\!/}
\newcommand{\Dslash}{D\!\!\!\!/}
\newcommand{\pslash}{p \hspace{-1.7mm} /}
\newcommand{\kslash}{k \hspace{-1.7mm} /}
\newcommand{\qslash}{q \hspace{-1.7mm} /}
\newcommand{\bs}{b \hspace{-1.7mm} /}


\date{}
\title{\textbf{On general ultraviolet properties of a class of confining propagators}}


 \author{\textbf{M.~A.~L.~Capri}\thanks{caprimarcio@gmail.com}\,\,
 \textbf{M.~S.~Guimaraes}\thanks{msguimaraes@uerj.br}\,\,,
 \textbf{I. Justo}\thanks{igorfjusto@gmail.com}\, \\ 
 \textbf{L.~F.~Palhares}\thanks{leticiapalhares@gmail.com}\,\,,
 \textbf{S.~P.~Sorella}\thanks{silvio.sorella@gmail.com}\,\,\,,\\[2mm]
 {\small \textnormal{  \it Departamento de F\'{\i }sica Te\'{o}rica, Instituto de F\'{\i }sica, UERJ - Universidade do Estado do Rio de Janeiro,}}
  \\ \small \textnormal{ \it Rua S\~{a}o Francisco Xavier 524, 20550-013 Maracan\~{a}, Rio de Janeiro, Brasil}\normalsize}

\maketitle

 
\begin{abstract}

We study the ultraviolet properties of theories whose fundamental fields display a confining, Gribov-type, propagator. These are propagators that exhibit complex poles and violate positivity, thus precluding a physical propagating particle interpretation. We show that the properties of this type of confining propagators do not change the ultraviolet behavior of the theory, in the sense that no new ultraviolet primitive divergences are generated, thus securing the renormalizability of these confining theories.  We illustrate these properties by studying a variety of models, including bosonic and fermionic confined degrees of freedom. The more intricate case of Super Yang-Mills with ${\cal N} =1$ supersymmetries in the Wess-Zumino gauge is taken as example in order to prove these statements to all orders by means of  the algebraic renormalization set up.

\end{abstract}

\section{Introduction}

The quantization of non-abelian gauge theories is still an open and rich subject for quantum field theorists.  A general framework  which takes into account the non-perturbative phenomena of gluon and quark confinement and of chiral symmetry breaking is still lacking. Needless to say, these issues represent a major challenge for our current understanding of non-abelian gauge theories in the non-perturbative infrared regime. 

A successful approach to investigate these topics within the context of the Euclidean quantum field theory is provided by the Gribov framework\footnote{For reviews on the Gribov issues, see \cite{Vandersickel:2012tz,Vandersickel:2011zc,Sobreiro:2005ec} and refs. therein.}. In his seminal work \cite{Gribov:1977wm},  Gribov pointed out that the Faddeev-Popov gauge-fixing procedure is plagued by the existence of Gribov copies. In a path integral formulation this manifests in the fact that the Faddeev-Popov operator develops zero modes. For example, in the Landau gauge, $\partial_\mu A^{a}_\mu=0$, the Faddeev-Popov operator is  given by ${\cal M}^{ab}=-(\partial^2 \delta^{ab} -g f^{abc}A^{c}_{\mu}\partial_{\mu})$. For sufficiently enough strong coupling constant, zero modes of this operator start to appear, rendering ill-defined  the  Faddeev-Propov procedure. To deal with this issue, Gribov  proposed to restrict the domain of integration in the path integral to a region in field space where the eigenvalues of the Faddeev-Popov operator are strictly positive. This  region is known as the Gribov region $\Omega$, being defined as 
\begin{align}
\Omega \;= \; \{ A^a_{\mu}\;; \;\; \partial_\mu A^a_{\mu}=0\;; \;\; {\cal M}^{ab}=-(\partial^2 \delta^{ab} -g f^{abc}A^{c}_{\mu}\partial_{\mu})\; >0 \; \} \;. \label{gr}
\end{align} 
It is useful to observe here that the inverse of the Faddeev-Popov operator, $({\cal M}^{-1})^{ab}$,  yields precisely the propagator ${\cal G}(k,A)$ of the Faddeev-Popov ghosts in the presence of an external gauge field $A^a_\mu$ \cite{Vandersickel:2012tz,Vandersickel:2011zc,Sobreiro:2005ec,Gribov:1977wm}, namely 
\begin{align}
 {\cal G}(k,A) = \frac{1}{(N^2-1)}\delta^{ab} \langle k | ({\cal M}^{-1})^{ab} | k \rangle   \;. \label{gpa}
\end{align}
This property can be employed to implement the restriction to the Gribov region $\Omega$ by requiring the absence of  poles in the ghost propagator for any non-zero value of the ghost external momentum $k$. This requirement  is known as the Gribov no-pole condition. More precisely, following \cite{Gribov:1977wm}, one can always represent the exact ghost propagator in the presence of an external gauge field as
\begin{align}
 {\cal G}(k,A) =  \frac{1}{k^2} (1+\sigma(k,A)) \;, \label{gp}
\end{align}
where $\sigma(k,A)$ is  the ghost form factor. Using the general properties of the diagrammatic expansion of  quantum field theory, we can write
\begin{align}
{\cal G} (k) = \langle {\cal G} (k,A)\rangle_{conn} = \frac{1}{k^2} (1 + \langle\sigma(k,A)\rangle_{conn}) =  \frac{1}{k^2} \frac{1}{(1 - \langle\sigma(k,A)\rangle_{1PI})}\label{np-gz}
\end{align}
where ``$conn$'' stands for the connected set of diagrams and $1PI$ denotes the 1-particle irreducible ones. It can be shown that $\langle\sigma(k,A)\rangle_{1PI}$ is a decreasing function of $k$ \cite{Vandersickel:2012tz,Vandersickel:2011zc,Sobreiro:2005ec,Gribov:1977wm}. Therefore, the condition that the ghost propagator has no  poles for any non-zero value of the ghost external momentum can be expressed as a condition for the maximum value of the ghost form factor, {\it i.e.}
\begin{align}
\langle\sigma(0,A)\rangle_{1PI} = 1\;.  \label{np-gz-cond}
\end{align}
Equation  \eqref{np-gz-cond} expresses the no-pole condition. In \cite{Capri:2012wx},  it has been shown that an exact closed expression for  $\sigma(0,A)$ can be obtained,  being proportional to the horizon function $H(A)$ of Zwanziger's  \cite{Vandersickel:2012tz} formalism: 
\begin{align}
\sigma(0, A) &=   -\frac{g^2}{V4(N^2-1)} \int \frac{d^4 p}{(2\pi)^4} \int \frac{d^4 q}{(2\pi)^4}  A^{ab}_{\mu_1}(-p) \left({\cal M}^{-1}\right)^{bc}_{pq}A^{ca}_{\mu}(q)=\frac{H(A)}{4V(N^2-1)}\,. \label{ff-hf}
\end{align}
The no-pole condition  \eqref{np-gz-cond} is thus equivalent to
\begin{align}
\langle H(A)\rangle_{1PI} =  V4(N^2-1) \;, 
 \label{np-gz-cond-3}
\end{align}
which is called the horizon condition. The relevance of the horizon function $H(A)$ relies on the fact that the restriction of the domain of integration in the functional integral to the Gribov region $\Omega$ can be effectively implemented by adding to the Yang-Mills action the quantity $H(A)$. More precisely, it turns out that, in the thermodynamic limit  \cite{Vandersickel:2012tz,Zwanziger:1988jt,Zwanziger:1989mf,Zwanziger:1992qr}, the partition function of the theory with the cut-off at the Gribov region $\Omega$ is given by  
\begin{align}
{\cal Z} = \; \int_{\Omega}  {\cal D}A\; \delta(\partial A)\; \det({\cal M}^{ab})\;  e^{-S_{YM}} = 
\int {\cal D}A\; \delta(\partial A)\; \det({\cal M}^{ab})\;  e^{-(S_{YM}+\gamma^4 H(A) - \gamma^4VD(N^2-1))} \;, \label{avac}
\end{align}
where the massive parameter $\gamma$ is known as the Gribov parameter \cite{Vandersickel:2012tz,Vandersickel:2011zc,Sobreiro:2005ec,Gribov:1977wm}. It is not a free parameter, being determined in a self-consistent way by the horizon condition  \eqref{np-gz-cond-3}, which can be rewritten as a stationary condition for the vacuum energy ${\cal E}$, {\it i.e.} 
\begin{align}
\frac{\partial {\cal E}}{\partial \gamma^2} = 0 \qquad  \Rightarrow   \qquad \langle H(A)\rangle_{1PI} =  V4(N^2-1) \;,  \label{stat}
\end{align}
where 
\begin{align}
{\cal Z} = e^{-{\cal E}}  \;. \label{vac}
\end{align}
Although expression \eqref{avac} is non-local, it can be fully localized by introducing a suitable set of auxiliary fields. The Faddeev-Popov measure is localized as usual by means of  the Faddeev-Popov ghosts $\left(c^a, \bar{c}^a\right)$ and of the Nakanishi-Lautrup field $b^a$. Moreover, the horizon function $H(A)$ can also be put in a local form  \cite{Vandersickel:2012tz,Zwanziger:1988jt,Zwanziger:1989mf,Zwanziger:1992qr} by using the auxiliary fields  $(\bar{\omega}_\mu^{ab}, \omega_\mu^{ab}, \bar{\varphi}_\mu^{ab},\varphi_\mu^{ab})$, where $(\bar{\varphi}_\mu^{ab},\varphi_\mu^{ab})$ are a pair of bosonic fields, while $(\bar{\omega}_\mu^{ab}, \omega_\mu^{ab})$ are anti-commuting. The resulting local action is called the Gribov-Zwanziger action $S_{GZ}$ \cite{Vandersickel:2012tz,Zwanziger:1988jt,Zwanziger:1989mf,Zwanziger:1992qr}, {\it i.e.}
\begin{align}
{\cal Z} = \int [D\phi] e^{-S_{GZ}}  \;, \label{vac}
\end{align}
where
\begin{equation} 
S_{GZ} = S_{FP}  + S_0+S_\gamma  \;, \label{sgz}
\end{equation}
with $S_{FP}$ being the Faddeev-Popov action in the Landau gauge 
\begin{align} 
S_{FP} = \frac{1}{4} \int d^{4}x \; F^{a}_{\mu \nu}F^{a}_{\mu\nu} + \int d^{4}x \left(  b^{a}\partial_{\mu}A^{a}_{\mu}
+\bar{c}^{a} \partial_{\mu}D^{ab}_{\mu}c^{b}  \right)  \;,    \label{FP}
\end{align}
and $S_0, S_{\gamma}$ given, respectively,  by 
\begin{align} 
S_0 &=\int d^{4}x \left( {\bar \varphi}^{ac}_{\mu} (\partial_\nu D^{ab}_{\nu} ) \varphi^{bc}_{\mu} - {\bar \omega}^{ac}_{\mu}  (\partial_\nu D^{ab}_{\nu} ) \omega^{bc}_{\mu}  - gf^{amb} (\partial_\nu  {\bar \omega}^{ac}_{\mu} ) (D^{mp}_{\nu}c^p) \varphi^{bc}_{\mu}  \right) \;, \label{s0}\\
S_\gamma &=\; \gamma^{2} \int d^{4}x \left( gf^{abc}A^{a}_{\mu}(\varphi^{bc}_{\mu} + {\bar \varphi}^{bc}_{\mu})\right)-4 \gamma^4V (N^2-1)\;. \label{hfl}
\end{align}
The GZ action \eqref{sgz} displays remarkable properties; it is renormalizable to all orders and has no extra free parameters with respect to the original Faddeev-Popov action \eqref{FP}; the parameter $\gamma$ is completely determined by the gap equation \eqref{stat} and does not renormalize independently. This means that the UV properties of the theory are not changed by restricting it to the Gribov region $\Omega$. Only two renormalization factors are in fact needed to renormalize the action  \eqref{sgz}  \cite{Vandersickel:2012tz,Zwanziger:1988jt,Zwanziger:1989mf,Zwanziger:1992qr}.

As discussed in  \cite{Dudal:2007cw,Dudal:2008sp,Dudal:2011gd,Baulieu:2008fy,Dudal:2009xh,Sorella:2009vt,Sorella:2010it,Capri:2010hb,Dudal:2012sb,Baulieu:2009ha,Cucchieri:2014via,Capri:2014fsa,Capri:2013naa,Capri:2011wp,Baulieu:2009xr}, the GZ action \eqref{sgz} breaks the standard BRST symmetry of the Faddeev-Popov action in a soft way, {\it i.e.} 
\begin{equation}
s S_{GZ} = \gamma^2 \Delta  \;, \label{brstbrr}
\end{equation}
where 
\begin{equation}
\Delta = \int d^{4}x \left( - gf^{abc} (D_\mu^{am}c^m) (\varphi^{bc}_{\mu} + {\bar \varphi}^{bc}_{\mu})   + g f^{abc}A^a_\mu \omega^{bc}_\mu            \right)  \;, \label{brstb1}
\end{equation}
and $s$ denotes the standard nilpotent BRST operator, defined by 
\begin{eqnarray}
\label{brst1}
sA^{a}_{\mu} &=& - D^{ab}_{\mu}c^{b}\;,\nonumber \\
s c^{a} &=& \frac{1}{2}gf^{abc}c^{b}c^{c} \;, \nonumber \\
s{\bar c}^{a} &=& b^{a}\;, \qquad \; \; 
sb^{a} = 0 \;, \nonumber \\
s{\bar \omega}^{ab}_\mu & = & {\bar \varphi}^{ab}_\mu \;, \qquad  s {\bar \varphi}^{ab}_\mu =0\;, \nonumber \\
s { \varphi}^{ab}_\mu&  = & {\omega}^{ab}_\mu  \;, \qquad s {\omega}^{ab}_\mu = 0 \;. 
\end{eqnarray}
We notice that the breaking term $\Delta$ is of dimension two in the quantum fields. As such, it is a soft breaking, which can be kept under control in the renormalization process \cite{Piguet:1995er}. 

Recently, it has been shown \cite{Capri:2015ixa} that a non-perturbative nilpotent extension of the standard BRST operator can be constructed in such a way that it is an exact symmetry of the GZ action. In particular, the existence of the soft breaking $\Delta$ turns out to be a consequence of the Ward identity stemming from the non-perturbative BRST exact symmetry \cite{Capri:2015ixa}. Nevertheless, as far as the UV renormalization of the GZ action is concerned, the standard softly broken BRST symmetry turns out to be very helpful. In fact, using the tools of the algebraic renormalization \cite{Piguet:1995er}, the softly broken identity \eqref{brstbrr} can be converted into useful Slavnov-Taylor identities which imply the all order UV renormalizability of expression \eqref{sgz}, see for example \cite{Dudal:2008sp,Capri:2014fsa,Capri:2013naa,Capri:2011wp,Baulieu:2009xr}.

We observe that,  in the local formulation, the horizon condition \eqref{stat} takes the form 
\begin{equation}
\langle g f^{abc} A^a_\mu(x) (\varphi^{bc}_{\mu}(x) + {\bar \varphi}^{bc}_{\mu}(x))  \rangle = 8 \gamma^2 (N^2-1) \;, \label{cndgz} 
\end{equation} 
which expresses the condensation of the local dimension two operator $g f^{abc} A^a_\mu (\varphi^{bc}_{\mu} + {\bar \varphi}^{bc}_{\mu})$. As shown in \cite{Dudal:2007cw,Dudal:2008sp,Dudal:2011gd}, this is not the only dimension two condensate present in the theory. The condensation of other dimension two operators, $A^a_\mu A^a_\mu$ and $\left( {\bar \varphi}^{ab}_{\mu}  \varphi^{ab}_{\mu} -{ \bar \omega}^{ab}_{\mu}  \omega^{ab}_{\mu} \right)$, turns out to be energetically favoured \cite{Dudal:2007cw,Dudal:2008sp,Dudal:2011gd}. The effective action which takes into account the formation of these condensates is known as the Refined-Gribov-Zwanziger (RGZ) action \cite{Dudal:2007cw,Dudal:2008sp,Dudal:2011gd}, being given by 
\begin{equation}
S_{RGZ} = S_{GZ} + \int d^4x \left(  \frac{m^2}{2} A^a_\mu A^a_\mu  - \mu^2 \left( {\bar \varphi}^{ab}_{\mu}  { \varphi}^{ab}_{\mu} -  {\bar \omega}^{ab}_{\mu}  { \omega}^{ab}_{\mu} \right)   \right)  \;,  \label{rgz}
\end{equation}
where the massive parameters $(m^2, \mu^2)$ are not independent and have a dynamical origin, being related to the existence of the dimension two condensates $\langle A^a_\mu A^a_\mu \rangle$ and  $\langle {\bar \varphi}^{ab}_{\mu}  { \varphi}^{ab}_{\mu} -  {\bar \omega}^{ab}_{\mu}  { \omega}^{ab}_{\mu}  \rangle$. As the GZ action, also the RGZ action can be proven to be tenormalizable to all orders  \cite{Dudal:2007cw,Dudal:2008sp,Dudal:2011gd}, while displaying the existence of a non-perturbative exact BRST symmetry  \cite{Capri:2015ixa}. 

The tree level gluon propagator obtained from the RGZ action \eqref{rgz}  reads
\begin{eqnarray} 
\langle  A^a_\mu(k)  A^b_\nu(-k) \rangle  & = &  \delta^{ab}  \left(\delta_{\mu\nu} - \frac{k_\mu k_\nu}{k^2}     \right)   {\cal D}(k^2) \;, \label{glrgz} \\
{\cal D}(k^2) & = & \frac{k^2 +\mu^2}{k^4 + (\mu^2+m^2)k^2 + 2Ng^2\gamma^4 + \mu^2 m^2}  \;. \label{Dg}
\end{eqnarray} 
It is worth mentioning that the infrared behaviour of the RGZ gluon propagator \eqref{glrgz}  and of the corresponding ghost two-point function turns out to be in remarkable agreement with the recent numerical lattice simulations  obtained on huge lattices  \cite{Cucchieri:2007rg,Cucchieri:2008fc,Cucchieri:2011ig}. Therefore, a numerical estimate of the non-perturbative parameters $(m,\mu,\gamma)$  can be obtained by fitting the lattice data by means of  expression \eqref{glrgz}, see \cite{Cucchieri:2011ig}. This leads to the presence of complex poles in the  gluon propagator \eqref{glrgz}, as well as to a violation of reflection positivity, precluding thus a physical particle interpretation. As a consequence, gluons cannot belong to the physical spectrum of the theory. We see thus that the restriction to the Gribov region $\Omega$ captures non-trivial aspects of the gluon confinement.  

Till now, the RGZ action has allowed for a variety of successful applications like: estimate of the masses of the first glueball states   \cite{Dudal:2010cd,Dudal:2013wja}, yielding results which display the right mass hierarchy as observed in the available numerical simulations and whose accuracy is comparable to other non-perturbative approaches to the glueball spectrum (cf. e.g. \cite{Mathieu:2008me} for a review), inclusion of quarks and estimate of the masses of meson states \cite{Dudal:2013vha}, study of the Casimir energy \cite{Canfora:2013zna}, finite temperature effects  \cite{Fukushima:2013xsa,Canfora:2013kma,Canfora:2015yia,Guimaraes:2015vra}, study of the confinement/deconfinement transition in presence of Higgs fields \cite{Capri:2012ah,Capri:2013oja}, analysis of the relevance of the Gribov issue in supersymmetric theories \cite{Capri:2014xea,Capri:2014tta}.

The feature that we want to explore in the present work is the fact that both the GZ and the RGZ tree-level propagators hold the key for the good UV behavior of the theory. More precisely we note that the propagator ${\cal D}(k^2)$ \eqref{Dg} can be written as 
\begin{eqnarray} 
{\cal D}(k^2) & = & \frac{k^2 +\mu^2}{k^4 + (\mu^2+m^2)k^2 + 2Ng^2\gamma^4 + \mu^2 m^2}  \;.\nonumber\\ 
& = &  \frac{1}{k^2 +  m^2} - \frac{2Ng^2\gamma^4}{\left(k^2 +  m^2\right)\left(k^2 +  M_{+}^2\right)\left(k^2 +  M_{-}^2\right)}
\label{Dg2}
\end{eqnarray}
where
\begin{eqnarray} 
M^2_{\pm}= \frac{\mu^2 + m^2}{2} \pm \frac 12 \sqrt{\left(\mu^2 + m^2 \right)^2 - 8Ng^2\gamma^4} 
\label{masses}
\end{eqnarray}
The first term in \eqref{Dg2} represents the usual propagator of a massive vector boson. The second term is the contribution coming from the restriction to the Gribov region. Notice the negative sign that points to an unphysical contribution that violates positivity requirements. The important feature we want to emphasize is the subleading contribution of the second term in the $UV$: it presents a $\sim 1/k^4$ suppression with respect to the standard first term, which will always produce a UV convergent loop contribution in dimension 4. The renormalization of the RGZ and GZ (corresponding to $\mu = m =0$) follows from this important property and, as already mentioned, it is well known that $\gamma$ does not renormalize independently and thus cannot be considered as an independent dynamically generated scale.

One is thus led to conjecture that this is a general property of theories displaying such confining propagators, with $\gamma$ standing for a general mass scale associated with confinement of the fundamental fields; $\gamma$ must be understood as a scale determined by other dynamically generated scales of the theory. More precisely, the second term in \eqref{Dg2} cannot generate any new $UV$ divergences in the theory and therefore cannot change the renormalization properties of the theory, which must be the same as with $\gamma = 0$. In a diagrammatic approach, only positive powers of propagators appear, so that it is clear that the highly-suppressed Gribov contribution (cf. \eqref{Dg2}, e.g.) will not influence the deep UV behavior of the theory. Furthermore, it follows that if the theory with $\gamma = 0$ does not generate a mass scale, then, since there can be no divergences proportional to $\gamma$, no mass scale will be generated in the  $\gamma \neq 0$ theory. This in turn means that it is not possible to assign a dynamical meaning to the parameter $\gamma$ in this case, {\it i.e.}, the only possible solution is to have  $\gamma = 0$ in these cases. An example of a theory displaying this feature is $N=4$ supersymmetric Yang-Mills which, due to its conformal character, has vanishing $\beta$-function to all orders. As a consequence of the absence of a scale, it turns out that $\gamma=0$, meaning that no mass scale associated to the Gribov copies is generated \cite{Capri:2014tta}.

In the following sections we will study a variety of examples that support these claims. In section \ref{confscalar} we discuss the case of an interacting scalar field theory displaying a confining propagator. In section 
\ref{confscalarferm} we consider the inclusion of confined fermions interacting with the confined scalars through a Yukawa term. In section \ref{symN1} we discuss the case of Super Yang-Mills with ${\cal N} =1$ supersymmetries
and show to all orders via the algebraic renormalization approach that the adoption of Gribov-type propagators does not produce any new UV divergences, with the renormalization of the IR parameters being completely defined by the UV renormalization of the parameters of the original theory. Section \ref{conc} collects our summary and conclusions.

\section{Interacting scalar fields with confining propagators}
\label{confscalar}

Consider the theory of a real scalar field $\phi$ defined by the following action in $D=4$ euclidean space
\begin{eqnarray} 
S = S_s + S_{int}+ S_{\gamma} 
\label{scalar}
\end{eqnarray}
where 
\begin{eqnarray} 
S_s  &=& \int d^4 x \left[ \frac 12 \phi \left( -\partial^2 + m^2 \right) \phi \right]\label{scalar2}\\
S_{int} &=&  \int d^4 x \; \left[\frac 14 \lambda \phi^4\right]\label{ints}\\
S_{\gamma} &=& \int d^4 x \left[ \frac 12 \phi \left(  \frac{\gamma^4}{-\partial^2}\right) \phi \right]\,,
\label{scalar3}
\end{eqnarray}
where $m^2$ is the mass of the scalar field in the deconfined ($\gamma\to 0$) theory and $\lambda$ is the quartic coupling. Here,  $\gamma$ is the confining parameter that shall play a similar role for the scalars as the Gribov mass does for the confined gluons. Our claim in this case is that the presence of the IR parameter $\gamma$ does not affect the deep UV behavior of the theory at all.

The quadratic part of the total action furnishes the tree-level confining propagator for the scalar fields:
\begin{eqnarray} 
{\cal D}(k^2) & = & \frac{k^2}{k^4 +m^2k^2 + \gamma^4}  \;.\nonumber\\ 
& = &  \frac{1}{k^2 +  m^2} - \frac{\gamma^4 }{\left(k^2 +  m^2\right)\left(k^2 +  M_{+}^2\right)\left(k^2 +  M_{-}^2\right)}
\nonumber\\ 
& = &  \frac{1}{k^2 +  m^2} -
\gamma^4 \Delta(k^2)
\label{scalarprop}
\end{eqnarray}
where we have isolated the confining contribution to the scalar propagator, $\gamma^4\Delta$, with
\begin{eqnarray} 
\Delta(k^2)&=& \frac{1}{\left(k^2 +  m^2\right)\left(k^2 +  M_{+}^2\right)\left(k^2 +  M_{-}^2\right)}\label{delta}\,,
\end{eqnarray}
which is highly suppressed in the UV: $\Delta\sim 1/k^6$.
The mass parameters $M^2_{\pm}$ are written in terms of $\gamma$ and $m$
\begin{eqnarray} 
M^2_{\pm}&=& \frac{m^2}{2} \pm \frac 12 \sqrt{m^4 - 4\gamma^4} \,,
\label{scalarmasses}
\end{eqnarray}
being complex for large enough $\gamma/m$. The complexity of these IR mass parameters is closely related to positivity violation and the absence of a physical particle interpretation for these excitations, in line with confinement.

It is straightforward to see that there are no new $UV$ divergences associated with the term $S_{\gamma}$ \eqref{scalar3} by looking at the diagrams of primitive divergences of the theory.

In fact, the one-loop scalar selfenergy is
\begin{eqnarray}
\Diagram{ 
&& c & \\
&&& \\
 f& & f &f
} \propto \int d^4 p{\cal D}(p) = \int d^4 p \frac{1}{p^2 +  m^2}  + \gamma^4 \int d^4p \Delta (p^2)\nonumber\\
 = \int d^4 p \frac{1}{p^2 +  m^2}  + \gamma^4 (\text{UV finite})
\end{eqnarray}

The correction to the quartic coupling at one loop reads:
\begin{eqnarray}
\Diagram{ 
fd \;\;\;\;\;\;\; & !{fl}{k-p} !{flu}{p}  & \;\;\;\;\;\;\; fu \\
fu \;\;\;\;\;\;\; &&  \;\;\;\;\;\;\; fd
} \propto \int d^4 p {\cal D}(k-p) {\cal D}(p) = \int d^4 p \frac{1}{p^2 +  m^2} \frac{1}{(k-p)^2 +  m^2} +\nonumber\\
&&\hspace{-10cm}+ \,\gamma^4 \int d^4p \Delta (p^2) \frac{1}{(k-p)^2 +  m^2} 
 + \gamma^4 \int d^4p  \frac{1}{p^2 +  m^2}\Delta ((k-p)^2) \nonumber \\
 &&\hspace{-10cm}+\, \gamma^8  \int d^4p \Delta (p^2) \Delta ((k-p)^2) \nonumber\\
= \int d^4 p \frac{1}{p^2 +  m^2} \frac{1}{(k-p)^2 +  m^2} + {\cal O}(\gamma^4, \gamma^8) (\text{UV finite }) 
\end{eqnarray}

As a representative example at two-loop order, we may look at the scalar selfenergy sunset diagram:
\begin{eqnarray}
\Diagram{ 
& !{fl}{k-p-q} !{flu}{p}  &\\
&& \\
f f & !{f}{q} f
} &\propto& \int d^4 p \int d^4 q {\cal D}(k-p-q) {\cal D}(q){\cal D}(p)\nonumber\\
&=&  \int d^4 p \int d^4 q \frac{1}{p^2 +  m^2} \frac{1}{q^2 +  m^2}\frac{1}{(k-p-q)^2 +  m^2} + \nonumber\\
 && + \,\gamma^4 \int d^4 p \int d^4 q  \Delta (p^2)  \frac{1}{q^2 +  m^2}\frac{1}{(k-p-q)^2 +  m^2} +\nonumber\\
&& + \,\gamma^4     \int d^4 p \int d^4 q \frac{1}{p^2 +  m^2} \Delta (q^2) \frac{1}{(k-p-q)^2 +  m^2}+\nonumber\\
 && + \,\gamma^4      \int d^4 p \int d^4 q \frac{1}{p^2 +  m^2} \frac{1}{q^2 +  m^2} \Delta ((k-p-q)^2) +{\cal O}(\gamma^ 8)\nonumber\\
&=&  \int d^4 p \frac{1}{p^2 +  m^2} \frac{1}{q^2 +  m^2}\frac{1}{(k-p-q)^2 +  m^2} + 
{\cal O}(\gamma^4, \gamma^8, \gamma^{12}) (\text{UV finite}) \nonumber\\
\end{eqnarray}

In all  examples above, the appearance of a general form for the contributions of the confining scale with increasingly UV convergent momentum integrals is clear.
It is straightforward to realize then that this pattern will spread throughout all orders of the diagrammatic expansion, so that
we are led 
to infer that  contributions proportional to $\gamma$  cannot give rise to new primitive divergences, besides the ones coming from $S_{s}$ \eqref{scalar2} alone, i.e. the original theory.

\section{Confined fermions and scalars with Yukawa interaction}
\label{confscalarferm}

The same reasoning can be applied when Dirac fermions are added to the theory, with an Yukawa coupling and a fermionic Gribov-type term rendering the fermionic excitations also confined.  

We consider here the theory in the absence of scalar condensates. In this case, the full action reads 
\begin{eqnarray} 
S = S_s + S_f+ S_{int}+ S_{\gamma, \Gamma} 
\label{scalarferm}
\end{eqnarray}
where 
\begin{eqnarray} 
S_s  &=& \int d^4 x \left( \frac 12 \phi \left( -\partial^2 + m^2 \right) \phi \right)\\
S_f &=&  \int d^4 x \;\bar{\psi} \left(\dslash + M\right) \psi \label{fermion}\\
S_{int} &=&  \int d^4 x \; \left(g\phi\bar{\psi} \psi +\frac 14 \lambda \phi^4\right)\label{int}\\
S_{\gamma, \Gamma} &=& \int d^4 x \left( \frac 12 \phi \left(  \frac{\gamma^4}{-\partial^2}\right) \phi  + \bar{\psi}\left(\frac{\Gamma^3}{-\partial^2}\right)\psi\right)
\label{scalar3F}
\end{eqnarray}
where $M$ is the mass of the original fermion field (i.e. for $\Gamma\to 0$) and $g$ is the Yukawa coupling. In the fermionic sector the IR mass scale analogous to the Gribov parameter is $\Gamma$. 

Analogously to the purely scalar case, it is easily seen that there are no $UV$ divergences associated to the whole term $S_{\gamma, \Gamma}$ \eqref{scalar3F}. 
The scalar excitations display the confining propagator of the last section,  \eqref{scalarprop}, while
for the confining fermion propagator, we have
\begin{eqnarray} 
{\cal S}(k^2) & = & \frac{i\kslash + M + \frac{\Gamma^3}{k^2}}{k^2 + (M + \frac{\Gamma^3}{k^2})^2}  \;.\nonumber\\ 
& = &  \frac{i\kslash + M }{k^2 + M^2} + \Gamma^3 \frac{(k^2+M^2)k^2 -(i\kslash + M)(2Mk^2 + \Gamma^3)  }{(k^6 + (M k^2 + \Gamma^3)^2)(k^2 + M^2)}
\nonumber\\ 
& = &  \frac{i\kslash + M }{k^2 + M^2} + \Gamma^3 \Sigma(k^2)\,,
\label{fermprop}
\end{eqnarray}
Again, the isolated confining contribution to the propagator is highly suppressed in the UV with respect to the standard massive Dirac term ($\sim 1/k$):
\begin{eqnarray} 
\Sigma(k)& = &   \frac{(k^2+M^2)k^2 -(i\kslash + M)(2Mk^2 + \Gamma^3)  }{(k^6 + (M k^2 + \Gamma^3)^2)(k^2 + M^2)}\sim 1/k^4
 \,,
\label{sigma}
\end{eqnarray}
and we anticipate that the primitive divergences of the theory with confined propagators will be exactly the ones coming from  $S_{s}$ \eqref{scalar2} and $S_{f}$ \eqref{fermion} alone,
since any contribution proportional to $\gamma$ or $\Gamma$ will be strongly suppressed in the UV.

At one loop order, besides the diagrams already analyzed in the previous section, new diagrams contributing to primitive divergences appear, due to the presence of fermion lines (dashed ones):
\begin{figure}[h!]
   \centering
       \includegraphics[width=0.8\linewidth]{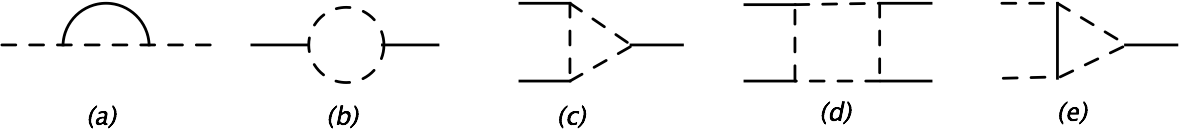}
               \caption{One-loop diagrams containing fermion (dashed) lines for the fermion and scalar selfenergies and cubic, quartic and Yukawa couplings, respectively. \label{fermiondiags}}
\end{figure}

It should be noticed that the Yukawa coupling breaks the discrete symmetry $\phi\mapsto -\phi$ originally present in the scalar sector, generating at the quantum level a cubic scalar interaction. This means that the renormalizable version of this theory requires a counterterm for the cubic scalar interaction, even if the physical value of this coupling is set to zero. In the case of a pseudoscalar Yukawa coupling (i.e. $g\phi\bar\psi\psi \mapsto g\phi\bar\psi\gamma^ 5\psi $), parity symmetry guarantees that the cubic terms vanish identically. We emphasize, however, that our statement concerning the UV properties of Gribov-type confining propagators remains valid in any case, as will be made explicit below via the whole set of primitive divergences at one loop order. 

In order to investigate the influence of the confining propagators in the UV regime, we may isolate the free fermion and scalar propagators from the confining contributions, namely $\Sigma(k)\stackrel{UV}{\sim} 1/k^4$ and $\Delta(k^2)\stackrel{UV}{\sim} 1/k^6$, being both highly suppressed in the UV. Writing down explicitly the momentum integrals in the corresponding expressions for the one-loop diagrams in Fig. \ref{fermiondiags}, we have, respectively:
\begin{enumerate}[label=(\alph*)]
\item the one-loop fermion self energy:
\begin{eqnarray}
 \int d^4 p {\cal D}(k-p) {\cal S}(p)
 &=&   \int d^4 p  \frac{1}{(k-p)^2 +  m^2} \frac{i\pslash +M}{p^2 +  M^2} +
\gamma^4  \int d^4 p {\Delta}((k-p)^2) \frac{i\pslash +M}{p^2 +  M^2}
\nonumber\\&&
+ \Gamma^3 \int d^4 p  \frac{1}{(k-p)^2 +  m^2} \Sigma(p)
+ \gamma^4\Gamma^3 \int d^4 p {\Delta}((k-p)^2)\Sigma(p) \nonumber
 \\
& =& \int d^4 p \frac{i\pslash +M}{p^2 +  M^2} \frac{1}{(k-p)^2 +  m^2} + {\cal O}(\gamma^4, \Gamma^3, \gamma^4\Gamma^3) (\text{UV finite}) 
\end{eqnarray}

\item  the fermion loop contributing to the scalar self energy:
\begin{equation}
 \int d^4 p\, {\rm Tr}[{\cal S}(p){\cal S}(k-p) ] = \int d^4 p {\rm Tr}\Big[\frac{i\pslash +M}{p^2 +  M^2} \frac{i(\kslash -\pslash)+ M}{(k-p)^2 +  M^2} \Big]+ {\cal O}( \Gamma^3, \Gamma^6) (\text{UV finite}) 
\end{equation}

\item the triangular diagram contributing to the scalar cubic interaction:
\begin{eqnarray}
 \int d^4 p\, {\rm Tr}[{\cal S}(p){\cal S}(p-k){\cal S}(p-k-k') ]
 &=& \int d^4 p {\rm Tr}\Big[ \frac{i\pslash +M}{p^2 +  M^2} \frac{i(\pslash-\kslash) +M}{(p-k)^2 +  M^2} \frac{i(\pslash -\kslash -\kslash')+ M}{(p-k-k')^2 +  M^2}\Big] +\nonumber \\
 &&+ {\cal O}( \Gamma^3, \Gamma^6, \Gamma^{9}) (\text{UV finite}) 
\end{eqnarray}

\item the fermion loop correction to the $\phi^4$ vertex:
\begin{eqnarray}
 \int d^4 p\,{\rm Tr}\Big[{\cal S}(p){\cal S}(p-k){\cal S}(p-k-k'){\cal S}(p-k-k'-k'')  \Big]
 &=& \nonumber\\&&\hspace{-10cm}=
  \int d^4 p {\rm Tr}\Big[ \frac{i\pslash +M}{p^2 +  M^2} \frac{i(\pslash-\kslash) +M}{(p-k)^2 +  M^2} \frac{i(\pslash -\kslash -\kslash')+ M}{(p-k-k')^2 +  M^2}
  \frac{i(\pslash -\kslash -\kslash'-\kslash'')+ M}{(p-k-k'-k'')^2 +  M^2}\Big] +\nonumber \\
 &&\hspace{-6cm}+ {\cal O}( \Gamma^3, \Gamma^6, \Gamma^{9},\Gamma^{12}) (\text{UV finite}) 
\end{eqnarray}

\item the modification of the Yukawa coupling:
\begin{eqnarray}
 \int d^4 p\,\Big[{\cal S}(p){\cal D}(p-k){\cal S}(p-k-k') \Big]
 &=& 
  \int d^4 p {\rm Tr}\Big[ \frac{i\pslash +M}{p^2 +  M^2}\frac{1}{(p-k)^2 +  m^2} \frac{i(\pslash-\kslash-\kslash') +M}{(p-k-k')^2 +  M^2} 
 \Big] +\nonumber \\
 &&+ {\cal O}( \gamma^4,\Gamma^3, \Gamma^6, \gamma^4\Gamma^3, \gamma^4\Gamma^6) (\text{UV finite}) 
\end{eqnarray}

\end{enumerate}

As already occurred for the confining scalar theory in the previous section, the highly suppressed UV behavior of the confining pieces $\Sigma(k)\stackrel{UV}{\sim} 1/k^4$ and $\Delta(k^2)\stackrel{UV}{\sim} 1/k^6$ enforces the convergence of all terms proportional to the new massive parameters introduced ($\gamma$ and $\Gamma$).
The divergent integrals in all diagrams above are exactly the ones coming from the original action, i.e. $S_s+S_f+S_{int}$ (cf. \eqref{scalarferm}). 
 In the theory including the confining quadratic nonlocal terms in $S_{\gamma,\Gamma}$, the absence of new primitive divergences then guarantees 
 that the parameters $\gamma$ and $\Gamma$ can be consistently related to dynamically generated scales and do not affect the UV regime of the theory.

Realizing that any diagrammatic expression at higher loops will involve higher powers of the propagators, it becomes straightforward to envision the generalization of our claim in the full diagrammatic expansion of this general Yukawa theory. Therefore, given the renormalizability of the original theory, one concludes that the resulting action with confining, Gribov-type propagators is renormalizable and the IR confining parameters in both fermionic and bosonic sectors do not display an independent renormalization, being thus consistent with dynamically generated mass scales.

\section{$\mathcal N=1$ Super Yang--Mills in Wess--Zumino gauge within the Gribov--Zwanziger approach}
\label{symN1}

Let us now investigate a more intricate theory with confining propagators, including gauge interactions as well as Majorana fermions. We consider here Yang-Mills theory in $D=4$ spacetime dimensions with $\mathcal N=1$ supersymmetry in the presence of the Gribov horizon. We shall use this (most complicated) example to prove, to all-orders in the loop expansion, our claim concerning the good UV behavior of Gribov-type propagators. The IR parameters introduced will be shown to have renormalization parameters that are completely determined by the renormalization of the original theory.

This theory has already been put forward and investigated in Ref. \cite{Capri:2014xea}. There, the extension of the Gribov-Zwanziger framework to $\mathcal  N = 1$ 
Super-Yang-Mills (SYM) theories quantized in the Wess-Zumino gauge by imposing the Landau gauge condition was presented. The resulting 
effective action is
\begin{equation} 
S_{SGZ}^{\mathcal{N}=1} = S_{SYM}^{\mathcal{N}=1}  + Q \int d^4x \left( {\check c}^a \partial_\mu A^a_\mu + {\tilde \omega}^{ac}_{\mu}  (-\partial_\nu D^{ab}_{\nu} ) \varphi^{bc}_{\mu}  \right) +  S_{\gamma} + S_{G} \;. \label{sgzn1}
\end{equation} 
Here, the operator $Q$ stands for the generalized BRST operator which encodes both gauge and supersymmetry transformations\footnote{For a detailed construction of  the operator $Q$ we refer to the Appendix \ref{algrenorm}. Notice also that the notation adopted here has few differences with respect to  the one employed  in the Introduction. In particular, the anti-ghost field is now denoted by $\check{c}^{a}$ instead of $\bar{c}^{a}$, while the fields $(\tilde\varphi^{ab}_{\mu},\tilde\omega^{ab}_{\mu})$ correspond  to the auxiliary Zwanziger fields $(\bar\varphi^{ab}_{\mu},\bar\omega^{ab}_{\mu})$, respectively. The meaning of this new notation is also clarified in Appendix \ref{algrenorm}.}; $S_{\gamma}$ is the horizon term in its local form, eq.\eqref{hfl}, namely 
\begin{equation}
S_\gamma =\; \gamma^{2} \int d^{4}x \left( gf^{abc}A^{a}_{\mu}(\varphi^{bc}_{\mu} + {\tilde \varphi}^{bc}_{\mu})\right)-4 \gamma^4V (N^2-1)\;; \label{hfln1}
\end{equation} 
and the term $S_{G}$ is given by 
\begin{equation}
S_{G} = - \frac{1}{2}M^3\int d^{4}x \left( \bar{\lambda}^{a\alpha}\frac{\delta_{\alpha\beta}}{\partial^{2}}\lambda^{a\beta}\right) \;, \label{sslambda}
 \end{equation} 
which also has a new massive constant $M$. This quantum action takes into account the existence of Gribov copies in the path-integral quantization of the theory. It encodes the restriction to the first Gribov horizon while
keeping full compatibility with non-perturbative supersymmetric features, such as the exactly vanishing vacuum energy. 

Even though this non-perturbative framework has been constructed through the introduction of two massive parameters $\gamma, M$ which are not present in the classical action,
those new parameters are determined in a dynamical, self-consistent way via two non-perturbative conditions: (i) the Gribov gap equation, that fixes $\gamma$ by imposing the positivity of the Faddeev-Popov operator and eliminating a large set of Gribov copies from the functional integral, and (ii) the vanishing of the vacuum energy, which determines the parameter $M$ that plays the role of a supersymmetric counterpart of the Gribov parameter $\gamma$, guaranteeing a consistent non-perturbative fermion sector. Interestingly, the appearance of the dynamical fermionic scale $M$ has been shown to be directly related to the formation of a gluino condensate, a well-known non-perturbative property of ${\cal N}=1$ SYM theories. For further details, the reader is referred to Ref. \cite{Capri:2014xea}. A brief summary of the notation adopted may also be found in the Appendix \ref{notations}.

The propagators of the theory \eqref{sgzn1} can be straightforwardly shown to be of the Gribov type. The gauge field propagator is:
\begin{equation}
\langle
A_{\mu}^a(p)A_{\nu}^b(-p)
\rangle
= \delta^{ab}\left(\delta_{\mu\nu}-\frac{p_{\mu}p_{\nu}}{p^2}\right)
\frac{p^2}{p^4+2Ng^2\gamma^4}\,,
\end{equation}
which, apart from the more complicated tensorial structure, is equivalent to the Gribov scalar propagator studied above in section \ref{confscalar}. The gauge field propagator in this Gribov-extended $\mathcal N=1$ SYM theory displays thus a confining contribution that is suppressed by an extra $1/p^4$ factor in the UV as compared to the free term.

For gluino fields we have:
\begin{eqnarray}
\langle
\bar{\lambda}_{\alpha}^a(p)
{\lambda}_{\beta}^b(-p)
\rangle
&=&
\frac{ip_{\mu}(\gamma_{\mu})_{\alpha\beta}+m(p^2)\delta_{\alpha\beta}}{p^2+m^2(p^2)} \; \delta^{ab}
\,,
\\
\langle
{\lambda}^{a \rho}(p)
{\lambda}_{\beta}^b(-p)
\rangle
&=&
- \frac{\left( ip_{\mu}(\gamma_{\mu})_{\alpha\beta}+m(p^2)\delta_{\alpha\beta}\right)}{p^2+m^2(p^2)}  \;\delta^{ab} C^{\alpha \rho}
\,,
\\
\langle
\bar{\lambda}^{a}_{\alpha}(p)
{\bar \lambda}^{b \tau}(-p)
\rangle
&=&
 \frac{\left( ip_{\mu}(\gamma_{\mu})_{\alpha\beta}+m(p^2)\delta_{\alpha\beta}\right)}{p^2+m^2(p^2)} \; \delta^{ab} C^{\beta \tau}
\,,
\end{eqnarray}
where $C^{\alpha\beta}$ is the charge conjugation matrix and
\begin{equation}
m(p^2)=\frac{M^3}{p^2}\,.
\end{equation}
The presence of three two-point correlation functions involving gluino fields is a result of the lack of charge conservation for Majorana fermions. One verifies however that all of them have the form of Gribov propagators with $M$ playing an analogous role as the Gribov parameter in the gluino sector. In particular, one can easily check that the same structure observed for the Gribov fermion propagator
in the previous section (cf. Eq.\eqref{fermprop}) is found here:
\begin{eqnarray} 
\langle
\bar{\lambda}_{\alpha}^a(k)
{\lambda}_{\beta}^b(-k)
\rangle& = & \frac{i\kslash + \frac{M^3}{k^2}}{k^2 + \frac{M^6}{k^4}} 
= \frac{i\kslash }{k^2} + M^3 \Sigma_{\lambda}(k^2)\,,
\label{gluinoprop}
\end{eqnarray}
where the isolated confining contribution $\Sigma_{\lambda}$ to the gluino propagator is again highly suppressed in the UV with respect to the leading term ($\sim 1/k$):
\begin{eqnarray} 
\Sigma_{\lambda}(k^2)& = &   \frac{k^4 -i\kslash M^3  }{(k^6 + M^6)k^2}\stackrel{\rm UV}{\sim} 1/k^4
 \,.
\label{sigmag}
\end{eqnarray}

The same reasoning applied in the scalar and Yukawa theories above may be followed here in order to prove that the UV regime of the theory remains the same even after the inclusion of nonlocal confining terms in the propagators. One may compute the one-loop primitive divergences and show that the confining parameters $\gamma,M$ will not affect the UV divergent pieces, due to the high suppression observed in the Gribov-type propagators. We shall, however, use this most complicated theory analyzed in the current section to present an all-order algebraic proof of renormalizability and of the fact that the confining parameters $\gamma,M$ do not display independent renormalization.

\noindent The non-local action \eqref{sgzn1} is, however, not helpful in the algebraic renormalization procedure. Fortunately we are able to write its local form with the insertion of auxiliary fields. 

\noindent The whole action which describes our model can then be written in its local form as, 
\begin{eqnarray}
S &=& S_{SYM} + S_{gf} + S_{GZ} + S^{local}_{{G}}\nonumber \\
&=&
\int d^{4}x\; \left\{\frac{1}{4}F^{a}_{\mu \nu}F^{a}_{\mu\nu} 
+ \frac{1}{2} \bar{\lambda}^{a\alpha} (\gamma_{\mu})_{\alpha\beta} D^{ab}_{\mu}\lambda^{b\beta}
+ \frac{1}{2}\mathfrak{D}^a\mathfrak{D}^a 
+ b^{a}\partial_{\mu}A^{a}_{\mu} \right. \nonumber\\
&&
+\check{c}^{a}\left[\partial_{\mu}D^{ab}_{\mu}c^{b}
-\bar{\epsilon}^{\alpha}(\gamma_{\mu})_{\alpha\beta}\partial_{\mu}\lambda^{a\,\beta}\right]
+ \tilde{\varphi}^{ac}_{\mu}\partial_{\nu}D_{\nu}^{ab}\varphi^{bc}_{\mu} 
-\tilde{\omega}^{ac}_{\mu}\partial_{\nu}D_{\nu}^{ab}\omega^{bc}_{\mu} \nonumber \\
&&
-gf^{abc}(\partial_{\nu}\tilde{\omega}^{ad}_{\mu})(D^{bk}_{\nu}c^{k})\varphi^{cd}_{\mu}
+gf^{abc}(\partial_{\nu}\tilde{\omega}^{ad}_{\mu})(\bar{\epsilon}^\alpha(\gamma_\nu)_{\alpha\beta}\lambda^{\beta b})\varphi^{cd}_{\mu} \nonumber \\
&&
+\gamma^{2}gf^{abc}A^{a}_{\mu}(\varphi^{bc}_{\mu} + \tilde{\varphi}^{bc}_{\mu}) 
-\gamma^{4}4(N_{c}^{2}-1) 
+\hat{\zeta}^{a\alpha} (\partial^{2} - \mu^{2})\zeta^{a}_{~\alpha} \nonumber \\
&&
\left.
-\hat{\theta}^{a\alpha}(\partial^{2} - \mu^{2})\theta^{a}_{~\alpha} 
-M^{3/2}(\bar{\lambda}^{a\alpha}\theta^{a}_{~\alpha} 
+\hat{\theta}^{a\alpha}\lambda^{a}_{~\alpha})
\right\}\;,
\label{thSYM}
\end{eqnarray} 
where the set of auxiliary fields $(\hat\theta^{a\alpha},\theta^{a\alpha},\hat\zeta^{a\alpha},\zeta^{a\alpha})$ has an analogous role of the set of auxiliary localizing fields $(\tilde\varphi^{a}_{\mu},\varphi^{a}_{\mu},\tilde\omega^{a}_{\mu},\omega^{a}_{\mu})$ introduced by Zwanziger in GZ model, i.e. it allows to describe the non-local gluino term \eqref{sslambda} in a local fashion.\\\\Applying the algebraic renormalization procedure to the local action \eqref{thSYM} above we are able to prove that:
(i) the Gribov-extended SYM theory is renormalizable; and (ii) the massive parameters $\gamma, M$ introduced in the infrared action do not renormalize independently, meaning that they are consistent with dynamically generated mass scales, produced by nonperturbative interactions in the original theory. All details of the proof may be found in the Appendix \ref{algrenorm}.

The final results for the renormalization factors related to the confining parameters $M,\gamma$ may be read from the renormalization of external sources conveniently introduced in the algebraic procedure (cf. Appendix \ref{algrenorm}). The renormalization of the sources $M$ and $\tilde{M}$ give us the renormalization factor of the Gribov parameter $\gamma^{2}$, while the renormalization of $V$ and $\hat{V}$ give us the renormalization of $M^{3/2}$, when every source assumes its physical value stated at \eqref{physval1}. We have:
\begin{eqnarray}
&&
Z_{\tilde{M}} =Z_{M} = Z_{\gamma^2}=  Z^{-1/2}_{g}Z^{-1/4}_{A}\;, \nonumber \\
&&
Z_{\hat{V}} =Z_{V} = Z_{M^{3/2}}=Z^{-1/2}_{\lambda}\;,
\end{eqnarray}
which clearly prove that the renormalization of the infrared parameters $M,\gamma$ is fixed by the renormalization factor of the original $SYM$ theory: the renormalization of the gauge coupling, $Z_g$, the wave function renormalization of the gauge field, $Z_A$, and and the wave function renormalization of the gluing field, $Z_{\lambda}$.

Therefore we conclude that this action is indeed a suitable nonperturbative infrared action for ${\cal N}=1$ SYM theories, reducing consistently to the ultraviolet original action. Moreover, even in this very intricate non-Abelian gauge theory with matter fields, the good UV behavior in the presence of confining propagators of the Gribov type shows up at all orders.

\section{Conclusion\label{conc}}

In this paper we have studied the UV behaviour of quantum field theory models in which the two-point correlation functions of the elementary fields are described by confining propagators of the Gribov type. 

Relying on the decompositions \eqref{scalarprop}, \eqref{fermprop}, we  have been able to show that the UV divergent behaviour of the Feynman diagrams is not affected by the infrared parameters, e.g. $(\gamma, \Gamma)$, encoded in the aforementioned confining propagators. 

From this property, it follows that no new UV divergences in the infrared parameters can arise. Otherwise said, the only UV divergences affecting the 1PI Green's functions of the theory are those present when the infrared parameters are set to zero. As a consequence, the infrared parameters do not renormalize independently, as explicitly shown in the case of $N=1$ supersymmetric Yang-Mills theory. 

In particular, in the case of a generic non-Abelian theory, the implementation of the restriction of the domain of integration to the Gribov region $\Omega$ has no consequences on the UV renormalisation properties of the theory. 

In conclusion, the main result of the present work can be  stated as follows: given a multiplicatively renormalizable 
Faddeev-Popov action, adding a Gribov horizon term in both gluon and matter sectors will not affect the ultraviolet properties of the theory at all. The resulting action remains multiplicatively renormalizable, with the same counterterms as the original theory. Moreover, the IR parameters originated by the horizon terms do not renormalize independently, being thus consistent with dynamically generated mass scales.

\section*{Acknowledgments}
The Conselho Nacional de Desenvolvimento Cient\'{\i}fico e
Tecnol\'{o}gico (CNPq-Brazil), the Faperj, Funda{\c{c}}{\~{a}}o de
Amparo {\`{a}} Pesquisa do Estado do Rio de Janeiro, the Latin
American Center for Physics (CLAF), the SR2-UERJ,  the
Coordena{\c{c}}{\~{a}}o de Aperfei{\c{c}}oamento de Pessoal de
N{\'{\i}}vel Superior (CAPES)  are gratefully acknowledged. L. F. P. 
is partially supported by a {\it Para Mulheres na Ci\^encia} grant 
from L'Or\'eal-UNESCO-ABC and by a BJT fellowship from the 
Brazilian program {\it Ci\^encia sem Fronteiras} (Grant No. 301111/2014-6).


\begin{appendix}

\section{Algebraic renormalization of $\mathcal  N=1$ Super Yang-Mills in Wess-Zumino gauge within the Gribov-Zwanziger approach} 
\label{algrenorm}
\subsection{Construction of a complete invariant action}
In order to prove the renormalizability of the action \eqref{thSYM} introduced in Sect.\ref{symN1},   we follow the procedure already employed in \cite{Zwanziger:1989mf,Dudal:2008sp,Capri:2013naa,Capri:2011wp,Baulieu:2009xr,Capri:2014xea} and embed the action \eqref{thSYM} into a more general one displaying a huge set of symmetries and Ward identities. In the present case, it turns out that the action \eqref{thSYM}  can be recovered as a particular case of the following expression  
\begin{equation}
\Sigma_{0}= S^{\mathcal{N}=1}_{SYM}+S_{gf}+S^{inv}_{GZ}+S^{conf}_{gluino}\,.
\label{Sigma_zero}
\end{equation}   
Let us proceed by specifying  the various terms appearing in the action \eqref{Sigma_zero}. The first term, $S^{\mathcal{N}=1}_{SYM}$, is the $\mathcal N=1$ Euclidean Super Yang-Mills action with Majorana fermions in the Wess-Zummino gauge, without matter fields, namely,  
\begin{equation}
\label{SYM}
S_\text{SYM} = \int d^{4}x \left[ \frac{1}{4}F^{a}_{\mu \nu}F^{a}_{\mu\nu} 
+ \frac{1}{2} \bar{\lambda}^{a\alpha} (\gamma_{\mu})_{\alpha\beta} D^{ab}_{\mu}\lambda^{b\beta}
+ \frac{1}{2}\mathfrak{D}^a\mathfrak{D}^a\right]\;,
\end{equation}
with  $D^{ab}_{\mu}=\delta^{ab}\partial_{\mu}-gf^{abc}A^{c}_{\mu}$ denoting the covariant derivative in the adjoint representation of the $SU(N)$; $\lambda_{\alpha}$ being a four component Majorana spinor; and  $\mathfrak{D}^{a}$ standing for a  dimension two auxiliary fields needed in order to close the algebra of ${\mathcal{N}=1}$ supersymmetry. Also,   $\bar\lambda=\lambda^{T}\mathcal{C}$ with $\mathcal{C}$ being the charge conjugation matrix, which is defined together with the Euclidean gamma matrices $\gamma_{\mu}$ in Appendix \ref{notations}. \\\\The second term in eq.\eqref{Sigma_zero} is the gauge fixing term in the  Landau gauge, given by 
\begin{equation}
S_{gf} = \int d^{4}x \left[\, \check{c}^{a}\left(\partial_{\mu}D^{ab}_{\mu}c^{b} - \bar{\epsilon}^{\alpha}(\gamma_{\mu})_{\alpha\beta}\,\partial_{\mu}\lambda^{a\,\beta} \,\right)+ b^{a}\,\partial_{\mu}A^{a}_{\mu} \,  \right]\;.
\label{gauge_fixing}
\end{equation}
In this term, $b^{a}$ is the  Lagrange multiplier enforcing the Landau gauge condition, $\partial_{\mu}A^{a}_{\mu}=0$, while $(c^{a}, \check{c}^{a})$ are the Fadeev-Popov ghost fields and $\bar{\epsilon}=\epsilon^{T}\mathcal{C}$ is a constant ghost. As we shall see later, the constant ghost  $\bar{\epsilon}$ is needed to encode the supersymmetric transformations into a unique generalized BRST operator.\\\\The third term of eq.\eqref{Sigma_zero} corresponds to the local and invariant Gribov-Zwanziger term\footnote{This term contains little modifications compared to the original non-supersymmetric formulation in order to accommodate both SUSY and BRST invariances.} being given by
\begin{eqnarray}
S^{inv}_{GZ}&=&\int d^{4}x\,\biggl\{\tilde{\varphi}^{ac}_{\mu}\partial_{\nu}D_{\nu}^{ab}\varphi^{bc}_{\mu} 
- \tilde{\omega}^{ac}_{\mu}\partial_{\nu}D_{\nu}^{ab}\omega^{bc}_{\mu} 
- gf^{abc}(\partial_{\nu}\tilde{\omega}^{ad}_{\mu})\left(D^{bk}_{\nu}c^{k}-\bar{\epsilon}^\alpha(\gamma_\nu)_{\alpha\beta}\lambda^{\beta b}\right)\varphi^{cd}_{\mu} 
\nonumber \\
&&
- N^{ab}_{\mu\nu}D^{ac}_{\mu}\tilde{\omega}^{cb}_{\nu} 
+ M^{ab}_{\mu\nu}\,\left[-D^{ac}_{\mu}\tilde{\varphi}^{cb}_{\nu} 
+ gf^{adc}\left(D^{dl}_{\mu}c^{l}-\bar{\epsilon}^{\alpha}(\gamma_{\mu})_{\alpha\beta}\lambda^{d\beta}\right)\tilde{\omega}^{cb}_{\nu} \right]
\nonumber \\
&&
- \tilde{M}^{ab}_{\mu\nu}D^{ac}_{\mu}\varphi^{cb}_{\nu} 
+ \tilde{N}^{ab}_{\mu\nu}\,\left[D^{ac}_{\mu}\omega^{cb}_{\nu} 
- gf^{adc}\left(D^{dl}_{\mu}c^{l}-\bar{\epsilon}^{\alpha}(\gamma_{\mu})_{\alpha\beta}\lambda^{d\beta}\right)\varphi^{cb}_{\nu} \right]
\nonumber \\
&&
- \tilde{M}^{ab}_{\mu\nu}M^{ab}_{\mu\nu} + \tilde{N}^{ab}_{\mu\nu}N^{ab}_{\mu\nu} \biggr\}\,.
\label{GribZwan}
\end{eqnarray}
In the expression above, $(\tilde{M}^{ab}_{\mu\nu},M^{ab}_{\mu\nu},\tilde{N}^{ab}_{\mu\nu},N^{ab}_{\mu\nu})$ are external sources which will be set equal to their physical values  after the renormalization procedure, {\it i.e.}  after removing the UV divergencies.\\\\Finally, the fourth and last term of the action \eqref{Sigma_zero} is  the local and invariant confining term for the gluino sector. This term is the analogous of the GZ term \eqref{GribZwan}.   It can be seen as the supersymmetric counterpart of expression \eqref{GribZwan}. It reads 
\begin{eqnarray}
S^{conf}_{gluino}&=&\int d^{4}x\,\biggl\{\hat{\zeta}^{a\alpha}\,\partial^{2}\zeta^{a}_{\alpha} 
- \hat{\theta}^{a\alpha}\,\partial^{2}\theta^{a}_{\alpha} 
- \hat{V}^{ab\,\alpha\beta}\,\bar{\lambda}^{a}_{\alpha}\theta^{b}_{\beta} 
+\hat{U}^{ab\,\alpha\beta}\biggl[- gf^{adc}c^{d}\bar{\lambda}^{c}_{\alpha}\theta^{b}_{\beta} 
\nonumber \\
&&
+ \frac{1}{2}\bar{\epsilon}^{\gamma}(\sigma_{\mu\nu})_{\gamma\alpha}F^{a}_{\mu\nu}\theta^{b}_{\beta} 
- \bar{\epsilon}^{\gamma}(\gamma_{5})_{\gamma\alpha}\mathfrak{D}^{a}\theta^{b}_{\beta}
+ \epsilon^{\gamma}(\gamma_{\mu})_{\gamma\eta}\bar{\epsilon}^{\eta}\bar{\lambda}^{a}_{\alpha}\partial_{\mu}\zeta^{b}_{\beta} \biggr]
- V^{ab\alpha\beta} \hat{\theta}^{b}_{\beta}\lambda^{a}_{\alpha}
\nonumber \\
&&
+ U^{ab\alpha\beta}\biggl[-\hat{\zeta}^{b}_{\beta}\lambda^{a}_{\alpha}
+ gf^{adc}\hat{\theta}^{b}_{\beta}c^{d}\lambda^{c}_{\alpha}
- \frac{1}{2}\hat{\theta}^{b}_{\beta} (\sigma_{\mu\nu})_{\alpha\gamma}\epsilon^{\gamma}F^{a}_{\mu\nu} 
+\hat{\theta}^{b}_{\beta}(\gamma_{5})_{\alpha\gamma}\epsilon^{\gamma}\mathfrak{D}^{a}\biggr]\biggr\}\,.
\label{gluino_term}
\end{eqnarray}
Analogously to the term \eqref{GribZwan}, this term depends on the external sources $(\hat{V}^{ab\,\alpha\beta},{V}^{ab\,\alpha\beta},\hat{U}^{ab\,\alpha\beta},{U}^{ab\,\alpha\beta})$.\\\\ As already mentioned, the original action \eqref{thSYM} can be re-obtained from expression \eqref{Sigma_zero} when the external sources attain the following physical values:
\begin{eqnarray}
\label{physval1}
&&
M^{ab}_{\mu\nu}\Big{|}_{phys}=\tilde{M}^{ab}_{\mu\nu}\Big{|}_{phys}=
\gamma^{2}\delta^{ab}\delta_{\mu\nu}\,,\qquad
N^{ab}_{\mu\nu}\Big{|}_{phys}=\tilde{N}^{ab}_{\mu\nu}\Big{|}_{phys}=0\,;
\nonumber \\
&&
V^{ab\alpha\beta}\Big{|}_{phys}=\hat{V}^{ab\alpha\beta}\Big{|}_{phys}=
-M^{3/2}\delta^{ab}\delta^{\alpha\beta}\,,\qquad
U^{ab\alpha\beta}\Big{|}_{phys}=\hat{U}^{ab\alpha\beta}\Big{|}_{phys}=0\,,
\end{eqnarray}
with
\begin{equation} 
	S = \Sigma_{0} \Big|_{phys}  \;. \label{ph}
\end{equation}
We see thus that  the action \eqref{thSYM} is a particular case of the more general expression \eqref{Sigma_zero}.  Therefore,  we will turn our attention to the action \eqref{Sigma_zero}, keeping in mind that we can always go back to the action \eqref{thSYM} by taking the limit \eqref{physval1}.\\\\The advantage of working with the most general action \eqref{Sigma_zero} is that it is left invariant by both  SUSY and BRST transformations, which can be embedded into a unique generalized BRST operator, see \cite{Capri:2014xea,Capri:2014jqa}, 
\begin{equation}
Q=s+\epsilon^{\alpha}\delta_{\alpha}\,,
\end{equation}
where $s$ is the usual BRST operator and  $\delta_{\alpha}$ are the SUSY generators, with $\epsilon^{\alpha}$ being the constant ghost.  More precisely, it turns out that  
\begin{equation}
Q\Sigma_0=0\,,
\end{equation} 
where the action of the operator $Q$ on the fields and external sources is defined as 
\begin{equation}
\left.
\begin{tabular}{rcll}
$QA^{a}_{\mu}$&$=$&$ - D^{ab}_{\mu}c^{b} 
+\bar{\epsilon}^\alpha(\gamma_\mu)_{\alpha\beta}\lambda^{a\beta}$&$\phantom{\Bigl|}$\cr
$Q\lambda^{a\alpha}$&$=$&$\displaystyle gf^{abc}c^{b}\lambda^{c\alpha}
- \frac{1}{2}(\sigma_{\mu\nu})^{\alpha\beta}\epsilon_{\beta} F_{\mu\nu}^{a}
+ (\gamma_{5})^{\alpha\beta}\epsilon_{\beta} \mathfrak{D}^a$&$\phantom{\Bigl|}$\cr
$Q\mathfrak{D}^a $&$=$&$gf^{abc}c^{b}\mathfrak{D}^c 
+ \bar{\epsilon}^{\alpha}(\gamma_{\mu})_{\alpha\beta}(\gamma_{5})^{\beta\eta}D_{\mu}^{ab}\lambda^{b}_{\eta} $&$\phantom{\Bigl|}$\cr
$Qc^{a}$&$=$&$\frac{1}{2}gf^{abc}c^{b}c^{c} 
- \bar{\epsilon}^{\alpha}(\gamma_{\mu})_{\alpha\beta}\epsilon^{\beta} A^{a}_{\mu}$&$\phantom{\Bigl|}$\cr
$Q\bar{c}^{a}$&$=$&$b^{a}$&$\phantom{\Bigl|}$\cr
$Qb^{a}$&$=$&$\nabla\bar{c}^{a}$&$\phantom{\Bigl|}$
\end{tabular}
\right\}\;,
\label{SBRST1}
\end{equation}
\begin{equation}
\left.
\begin{tabular}{rcll}
$Q\varphi^{ab}_{\mu}$&$=$&$\omega^{ab}_{\mu}$&$\phantom{\Bigl|}$\cr
$Q\omega^{ab}_{\mu}$&$=$&$\nabla\varphi^{ab}_{\mu}$&$\phantom{\Bigl|}$\cr
$Q\tilde\omega^{ab}_{\mu}$&$=$&$\tilde\varphi^{ab}_{\mu}$&$\phantom{\Bigl|}$\cr
$Q\tilde\varphi^{ab}_{\mu}$&$=$&$\nabla\tilde\omega^{ab}_{\mu}$&$\phantom{\Bigl|}$\cr
\end{tabular}
\right\}\;,\qquad
\left.
\begin{tabular}{rcll}
$QM^{ab}_{\mu\nu}$&$=$&$N^{ab}_{\mu\nu}$&$\phantom{\Bigl|}$\cr
$QN^{ab}_{\mu\nu}$&$=$&$\nabla M^{ab}_{\mu\nu}$&$\phantom{\Bigl|}$\cr
$Q\tilde{N}^{ab}_{\mu\nu}$&$=$&$\tilde{M}^{ab}_{\mu\nu}$&$\phantom{\Bigl|}$\cr
$Q\tilde{M}^{ab}_{\mu\nu}$&$=$&$\nabla\tilde{N}^{ab}_{\mu\nu}$&$\phantom{\Bigl|}$\cr
\end{tabular}
\right\}\;,
\label{SBRST2}
\end{equation}
\begin{equation}
\left.
\begin{tabular}{rcll}
$Q\zeta^{a}_{\alpha}$&$=$&$\theta^{a}_{\alpha}$&$\phantom{\Bigl|}$\cr
$Q\theta^{a}_{\alpha}$&$=$&$\nabla\zeta^{a}_{\alpha}$&$\phantom{\Bigl|}$\cr
$Q\hat\theta^{a}_{\alpha}$&$=$&$\hat\zeta^{a}_{\alpha}$&$\phantom{\Bigl|}$\cr
$Q\hat\zeta^{a}_{\alpha}$&$=$&$\nabla\hat\theta^{a}_{\alpha}$&$\phantom{\Bigl|}$\cr
\end{tabular}
\right\}\;,\qquad
\left.
\begin{tabular}{rcll}
$QU^{ab\,\alpha\beta}$&$=$&$V^{ab\,\alpha\beta}$&$\phantom{\Bigl|}$\cr
$QV^{ab\,\alpha\beta}$&$=$&$\nabla U^{ab\,\alpha\beta}$&$\phantom{\Bigl|}$\cr
$Q\hat{U}^{ab\,\alpha\beta}$&$=$&$\hat{V}^{ab\,\alpha\beta}$&$\phantom{\Bigl|}$\cr
$Q\hat{V}^{ab\,\alpha\beta}$&$=$&$\nabla \hat{U}^{ab\,\alpha\beta}$&$\phantom{\Bigl|}$\cr\end{tabular}
\right\}\;.
\label{SBRST3}
\end{equation}
The operator $Q$ has the pleasant property that its square gives the generator of the translations  \cite{Capri:2014xea,Capri:2014jqa}, {\it i.e.}  where $Q^{2}$ is  defined as  the translation operator 
\begin{equation}
\label{top}
Q^{2}\equiv\nabla = \bar{\epsilon}^{\alpha}(\gamma_{\mu})_{\alpha\beta}\epsilon^{\beta} \partial_{\mu}\,.
\end{equation}
Notice that the transformations $QA^{a}_{\mu}$, $Q\lambda^{a\beta}$, $QD^{a}$ and $Qc^{a}$  are non-linear in the quantum fields, meaning that  they have to be treated as  composite operators.  Therefore, following the algebraic renormalization procedure \cite{Piguet:1995er}, we introduce them into the starting action  coupled to suitable external sources $(K^a_\mu, L^a,T^a, Y^{a \alpha})$ :\begin{equation}
S_{sources}=\int d^{4}x \left[ -Q(K^{a}_{\mu} A^{a}_{\mu}) + Q(L^{a}c^{a}) - Q(T^{a}\mathfrak{D}^{a}) + Q(Y^{a\alpha}\lambda^{a}_{\alpha}) \right]\;,
\end{equation}
with the following transformations
\begin{equation}
\left.\begin{tabular}{rcl}
$QK^{a}_{\mu}$&$\!\!\!\!=\!\!\!\!$&$\Omega^{a}_{\mu}\phantom{\Bigl|}$\cr
$Q\Omega^{a}_{\mu}$&$\!\!\!\!=\!\!\!\!$&$\nabla K^{a}_{\mu}\phantom{\Bigl|}$
\end{tabular}\right\}\,,\quad
\left.\begin{tabular}{rcl}
$QL^{a}$&$\!\!\!\!=\!\!\!\!$&$\Lambda^{a}\phantom{\Bigl|}$\cr
$Q\Lambda^{a}$&$\!\!\!\!=\!\!\!\!$&$\nabla L^{a}\phantom{\Bigl|}$
\end{tabular}\right\}\,,\quad
\left.\begin{tabular}{rcl}
$QT^{a}$&$\!\!\!\!=\!\!\!\!$&$J^{a}\phantom{\Bigl|}$\cr
$QJ^{a}$&$\!\!\!\!=\!\!\!\!$&$\nabla T^{a}\phantom{\Bigl|}$
\end{tabular}\right\}\,,\quad
\left.\begin{tabular}{rcl}
$QY^{a\alpha}$&$\!\!\!\!=\!\!\!\!$&$X^{a\alpha}\phantom{\Bigl|}$\cr
$QX^{a\alpha}$&$\!\!\!\!=\!\!\!\!$&$\nabla Y^{a\alpha}\phantom{\Bigl|}$
\end{tabular}\right\}\,.
\end{equation}
Finally, for the complete starting invariant action suitable to study the symmetry content and  renormalizability of the theory, we have  
\begin{eqnarray}
\label{fullact}
\Sigma &=& \Sigma_{0}+S_{sources}\nonumber\\\cr
&=&\int d^{4}x \biggl\{ 
\frac{1}{4}F^{a}_{\mu \nu}F^{a}_{\mu\nu} 
+ \frac{1}{2}\bar{\lambda}^{a\,\alpha} (\gamma_{\mu})_{\alpha\beta}\,D^{ab}_{\mu}\lambda^{b\,\beta}
+ \frac{1}{2}\mathfrak{D}^{a}\mathfrak{D}^{a}
+ b^{a}\partial_{\mu}A^{a}_{\mu}
+ \check{c}^{a}\Bigl[\partial_{\mu}D^{ab}_{\mu}c^{b}
- \bar{\epsilon}^{\alpha}(\gamma_{\mu})_{\alpha\beta}\partial_{\mu}\lambda^{a\,\beta}\Bigr]\nonumber \\\cr
&&
+\tilde{\varphi}^{a}_{i}\partial_{\mu}D_{\mu}^{ab}\varphi^{b}_{i}
-\tilde{\omega}^{a}_{i}\partial_{\mu}D_{\mu}^{ab}\omega^{b}_{i}
-gf^{abc}(\partial_{\mu}\tilde{\omega}^{a}_{i})\Bigl[(D^{bd}_{\mu}c^{d})
-\bar{\epsilon}^\alpha(\gamma_\mu)_{\alpha\beta}\lambda^{b\beta}\Bigl]
\varphi^{c}_{i}\nonumber\\\cr
&&
-{N}^{a}_{\mu{i}}\,D^{ab}_{\mu}\tilde{\omega}^{b}_{i}
-{M}^{a}_{\mu{i}}\Bigl[D^{ab}_{\mu}\tilde{\varphi}^{b}_{i}
-gf^{abc}(D^{bd}_{\mu}c^{d})\tilde{\omega}^{c}_{i}
+gf^{abc}\bar{\epsilon}^{\alpha}(\gamma_{\mu})_{\alpha\beta}\lambda^{b\beta}\tilde{\omega}^{c}_{i}\Bigr]\nonumber \\\cr
&&
-\tilde{M}^{a}_{\mu{i}}\,D^{ab}_{\mu}\varphi^{b}_{i}
+\tilde{N}^{a}_{\mu{i}}\Bigl[D^{ab}_{\mu}\omega^{b}_{i}
- gf^{abc}(D^{bd}_{\mu}c^{d})\varphi^{c}_{i}
+ gf^{abc}\bar{\epsilon}^{\alpha}(\gamma_{\mu})_{\alpha\beta}\lambda^{b\beta}\varphi^{c}_{i}\Bigr]\nonumber \\\cr
&&
- \tilde{M}^{a}_{\mu{i}}{M}^{a}_{\mu{i}}
+ \tilde{N}^{a}_{\mu{i}}{N}^{a}_{\mu{i}}
+ \hat{\zeta}^{I}\,\partial^{2}\zeta_{I}
- \hat{\theta}^{I}\,\partial^{2}\theta_{I}
+ \hat{V}^{Ia\alpha}\,\bar{\lambda}^{a}_{\alpha}\theta_{I}\nonumber \\\cr
&&
- \hat{U}^{Ia\alpha}\Bigl[gf^{abc}c^{b}\bar{\lambda}^{c}_{\alpha}\theta_{I}
-\bar{\lambda}^{a}_{\alpha}\nabla\zeta_{I}
-\frac{1}{2}\bar{\epsilon}^{\gamma}(\sigma_{\mu\nu})_{\gamma\alpha}F^{a}_{\mu\nu}\theta_{I} 
+\bar{\epsilon}^{\gamma}(\gamma_{5})_{\gamma\alpha}\mathfrak{D}^{a}\theta_{I}\Bigr]\nonumber \\\cr
&&
+ V^{Ia\alpha}\,\hat{\theta}_{I}\lambda^{a}_{\alpha}
+ U^{Ia\alpha}\Bigl[-\hat{\zeta}_{I}\lambda^{a}_{\alpha}
+ gf^{abc}\hat{\theta}_{I}c^{b}\lambda^{c}_{\alpha}
- \frac{1}{2}\hat{\theta}_{I}(\sigma_{\mu\nu})_{\alpha\gamma}\epsilon^{\gamma}F^{a}_{\mu\nu}  
+ \hat{\theta}_{I}(\gamma_{5})_{\alpha\gamma}\epsilon^{\gamma}\mathfrak{D}^{a}\Bigr] \nonumber \\\cr
&&
- \Omega^{a}_{\mu} A^{a}_{\mu}
- K^{a}_{\mu}\Bigl[D^{ab}_{\mu}c^{b} - \bar{\epsilon}^\alpha(\gamma_\mu)_{\alpha\beta}\lambda^{a\beta}\Bigr]
+ \Lambda^{a}c^{a}
+ L^{a}\Bigl[ \frac{g}{2}f^{abc}c^{b}c^{c}
- \bar{\epsilon}^\alpha(\gamma_{\mu})_{\alpha\beta}\epsilon^\beta A^{a}_{\mu}\Bigr]\nonumber \\\cr
&&
- J^{a}\mathfrak{D}^{a}
+ T^{a}\Bigl[gf^{abc}c^{b}\mathfrak{D}^{c}
+ \bar{\epsilon}^{\alpha}(\gamma_{\mu})_{\alpha \beta}(\gamma_{5})^{\beta\eta}D_{\mu}^{ab}\lambda^{b}_{\eta}\Bigr]\nonumber\\\cr
&&+ X^{a\alpha}\lambda^{a}_{\alpha}
+Y^{a\alpha}\Bigl[ gf^{abc}c^{b}\lambda^{c}_{\alpha} - \frac{1}{2}(\sigma_{\mu\nu})_{\alpha\beta} F_{\mu\nu}^{a}\epsilon^{\beta}
+ (\gamma_{5})_{\alpha\beta}\epsilon^{\beta} \mathfrak{D}^a \Bigr]
\biggr\}\;,
\end{eqnarray}
where, following \cite{Zwanziger:1989mf,Dudal:2008sp,Capri:2013naa,Capri:2011wp,Baulieu:2009xr,Capri:2014xea},  we have introduced the composite index notation:
\begin{eqnarray}
\left(\varphi^{ab}_{\mu},\tilde\varphi^{ab}_{\mu}, \omega^{ab}_{\mu},\tilde\omega^{ab}_{\mu}\right)&\to&
\left(\varphi^{a}_{i},\tilde\varphi^{a}_{i}, \omega^{a}_{i},\tilde\omega^{a}_{i}\right)\,,\nonumber\\
\left(M^{ab}_{\mu\nu},\tilde{M}^{ab}_{\mu\nu},N^{ab}_{\mu\nu},\tilde{N}^{ab}_{\mu\nu}\right)&\to&
\left(M^{a}_{\mu i},\tilde{M}^{a}_{\mu i},N^{a}_{\mu i},\tilde{N}^{a}_{\mu i}\right)
\end{eqnarray}
and
\begin{eqnarray}
\left(\theta^{a}_{\alpha},\hat\theta^{a}_{\alpha}, \zeta^{a}_{\alpha},\hat\zeta^{a}_{\alpha}\right)&\to&
\left(\theta_{I},\hat\theta_{I}, \zeta_{I},\hat\zeta_{I}\right)\,,\nonumber\\
\left(V^{ab\,\alpha\beta},\hat{V}^{ab\,\alpha\beta},U^{ab\,\alpha\beta},\hat{U}^{ab\,\alpha\beta}\right)&\to&
\left(V^{Ia\alpha},\hat{V}^{Ia\alpha},U^{Ia\alpha},\hat{U}^{Ia\alpha}\right)\,,
\end{eqnarray}
with
\begin{eqnarray}
(a,\mu)&\equiv& i,j,k,l,\dots\in\{1,\dots,4(N^{2}-1)\}\,,\nonumber\\
(a,\alpha)&\equiv& I,J,K,L,\dots\in\{1,\dots,4(N^{2}-1)\}\,.
\end{eqnarray}
A summary of all indices used here can be found in Appendix \ref{notations},  where  we also display the quantum numbers,  {\it i.e.} mass dimensions and charges of all fields and sources of the model, see  Tables \ref{tabfields} and \ref{tabsources}.    According to those Tables, the fields and sources denoted with a  $(\,\tilde\/\,)$ have charge $q_f$ equal to $(-1)$, while the fields and sources denoted with a $(\,\hat\/\,)$ have charge $q_{f'}$ equal to $(-1)$. That is why in Sect.\ref{symN1}, a slightly different notation with respect to that used in  the Introduction has been adopted.
\subsection{Symmetry content of the model}
It turns out that the the complete action  \eqref{fullact} displays a huge set of Ward identities, which we enlist below:  
\subsubsection{Ward identities}
\begin{itemize}
{\item The Slavnov-Taylor identity:}
\begin{equation}
\mathcal{S}(\Sigma) = 0 \;, \label{sti}
\end{equation}
with
\begin{eqnarray}
\mathcal{S}(\Sigma) &\equiv& \int d^{4}x \biggl\{\biggl(\frac{\delta \Sigma}{\delta A^{a}_{\mu}}
+ \Omega^{a}_{\mu}\biggr)\frac{\delta \Sigma}{\delta K^{a}_{\mu}}
+ \biggl(\frac{\delta \Sigma}{\delta \lambda^{a\alpha}}
+ X^{a\alpha}\biggr)\frac{\delta \Sigma}{\delta Y^{a\alpha}}
+\biggl(\frac{\delta \Sigma}{\delta c^{a}}
+ \Lambda^{a}\biggr)\frac{\delta \Sigma}{\delta L^{a}} \nonumber \\
&&
+ \biggl(\frac{\delta \Sigma}{\delta \mathfrak{D}^{a}}
+ {J}^{a}\biggr)\frac{\delta \Sigma}{\delta T^{a}}
+ b^{a}\frac{\delta \Sigma}{\delta \check{c}^{a}}
+ \omega^{a}_{i}\frac{\delta \Sigma}{\delta \varphi^{a}_{i}}
+ \tilde{\varphi}^{a}_{i}\frac{\delta \Sigma}{\delta \tilde{\omega}^{a}_{i}}
+ \hat{\zeta}^{I}\frac{\delta \Sigma}{\delta \hat{\theta}^{I}}
+ \theta^{I}\frac{\delta \Sigma}{\delta \zeta^{I}}\nonumber \\
&&
+{V}^{Ia\alpha}\frac{\delta \Sigma}{\delta {U}^{Ia\alpha}}
+ \hat{V}^{Ia\alpha}\frac{\delta \Sigma}{\delta \hat{U}^{Ia\alpha}}
+ {N}^{a}_{\mu i}\frac{\delta \Sigma}{\delta{M}^{a}_{\mu i}}
+ \tilde{M}^{a}_{\mu i}\frac{\delta \Sigma}{\delta \tilde{N}^{a}_{\mu i}} \nonumber  \\
&&
+(\nabla{U}^{Ia\alpha})\frac{\delta \Sigma}{\delta{V}^{Ia\alpha}}
+(\nabla\hat{U}^{Ia\alpha})\frac{\delta \Sigma}{\delta \hat{V}^{Ia\alpha}}
+(\nabla{M}^{a}_{\mu i})\frac{\delta \Sigma}{\delta {N}^{a}_{\mu i}} \nonumber \\
&&
+(\nabla\tilde{N}^{a}_{\mu i})\frac{\delta \Sigma}{\delta \tilde{M}^{a}_{\mu i}}
+(\nabla K^{a}_{\mu})\frac{\delta \Sigma}{\delta \Omega^{a}_{\mu}}
+(\nabla Y^{a\alpha})\frac{\delta \Sigma}{\delta {X}^{a\alpha}}
+(\nabla T^{a})\frac{\delta \Sigma}{\delta {J}^{a}}
+(\nabla L^{a})\frac{\delta \Sigma}{\delta {\Lambda}^{a}}\nonumber  \\
&&
+(\nabla \check{c}^{a})\frac{\delta \Sigma}{\delta b^{a}}
+(\nabla\varphi^{a}_{i})\frac{\delta \Sigma}{\delta \omega^{a}_{i}}
+(\nabla\tilde{\omega}^{a}_{i})\frac{\delta \Sigma}{\delta \tilde{\varphi}^{a}_{i}}
+(\nabla\hat{\theta}^{I})\frac{\delta \Sigma}{\delta \hat{\zeta}^{I}}
+(\nabla\zeta^{I})\frac{\delta \Sigma}{\delta \theta^{I}}  \biggr\}
\,.
\label{ST}
\end{eqnarray}
\item{The gauge-fixing condition and anti-ghost equation:}
\begin{equation}
\frac{\delta\Sigma}{\delta b^{a}}=\partial_{\mu}A^{a}_{\mu}\,,\qquad
\check{G}^{a}(\Sigma)\equiv\left(\frac{\delta}{\delta\check{c}^{a}}+\partial_{\mu}\frac{\delta}{\delta K^{a}_{\mu}}\right)\Sigma=0\,.
\label{GFandAntiGhost}
\end{equation}
\item{The equations of motion of the auxiliary fields:}
\begin{eqnarray}
\widetilde{\mathcal{F}}^{a}_{i}(\Sigma)&\!\!\!\equiv\!\!\!&\left(\frac{\delta}{\delta \tilde{\varphi}^{a}_{i}} + \partial_{\mu}\frac{\delta}{\delta \tilde{M}^{a}_{\mu i}} - gf^{abc}M^{b}_{\mu i}\frac{\delta}{\delta\Omega^{c}_{\mu}}\right)\Sigma = 0 \,,
\\\cr
\mathcal{W}^{a}_{i}(\Sigma)&\!\!\!\equiv\!\!\!&\left[\frac{\delta}{\delta\omega^{a}_{i}} + \partial_{\mu}\frac{\delta}{\delta N^{a}_{\mu i}} - gf^{abc} \left( \frac{\delta }{\delta b^{c}} \tilde{\omega}^{b}_{i} + \frac{\delta }{\delta \Omega^{c}_{\mu}}\tilde{N}^{b}_{\mu i} \right)\right]\Sigma= 0 \,,
\\\cr
\widetilde{\mathcal{W}}^{a}_{i}(\Sigma)&\!\!\!\equiv\!\!\!&\left[\frac{\delta }{\delta \tilde{\omega}^{a}_{i}} 
+ \partial_{\mu}\frac{\delta}{\delta\tilde{N}^{a}_{\mu i}} 
- gf^{abc}\left( M^{b}_{\mu i}\frac{\delta}{\delta K^{c}_{\mu}} 
- N^{b}_{\mu i} \frac{\delta}{\delta\Omega^{c}_{\mu}} \right)\right]\Sigma=
0 \,,
\\\cr
\mathcal{F}^{a}_{i}(\Sigma)&\!\!\!\equiv\!\!\!&
\left[\frac{\delta}{\delta\varphi^{a}_{i}} 
+ \partial_{\mu}\frac{\delta}{\delta M^{a}_{\mu i}} 
- gf^{abc}\left( \frac{\delta }{\delta b^{c}}\tilde{\varphi}^{b}_{i} 
+ \frac{\delta }{\delta \Omega^{c}_{\mu}}\tilde{M}^{b}_{\mu i} 
+ \frac{\delta }{\delta \check{c}^{b}} \tilde{\omega}^{c}_{i} 
- \tilde{N}^{c}_{\mu i} \frac{\delta }{\delta K^{b}_{\mu}} \right)\right]\Sigma = 0\,,
\\\cr
\widehat{\mathcal{T}}^{I}(\Sigma)&\!\!\!\equiv\!\!\!&\left(\frac{\delta}{\delta\hat\theta_{I}}
-{U}^{Ia\alpha}\,\frac{\delta}{\delta Y^{a\alpha}}\right)\Sigma=-\partial^{2}{\theta}^{I}
+{V}^{Ia\alpha}\,\lambda^{a}_{\alpha}\,,
\\\cr
\mathcal{T}^{I}(\Sigma)&\!\!\!\equiv\!\!\!&\left[\frac{\delta}{\delta\theta_{I}}
-\left(\frac{\delta}{\delta Y^{a}}\right)^{T}_{\beta}\mathcal{C}^{\beta}_{\;\;\alpha}\;{\hat{U}}^{Ia\alpha}\right]\Sigma=\partial^{2}{\hat\theta}^{I}
-\bar\lambda^{a}_{\;\;\alpha}\;{\hat{V}}^{Ia\alpha}\,,
\\\cr
\frac{\delta\Sigma}{\delta\zeta_{I}}&\!\!\!=\!\!\!&
\partial^{2}\hat{\zeta}^{I}
- \nabla(\hat{U}^{Ia\alpha}\,\bar\lambda^{a}_{\alpha})\,,
\\\cr
\frac{\delta\Sigma}{\delta\hat\zeta_{I}}&\!\!\!=\!\!\!&\partial^{2}{\zeta}^{I}
-{U}^{Ia\alpha}\,\lambda^{a}_{\alpha}\,,
\\\cr
\frac{\delta\Sigma}{\delta{\mathfrak{D}^{a}}}&\!\!\!=\!\!\!&-\mathfrak{D}^{a}- J^{a} + gf^{abc}c^{b}T^{c}-Y^{a\alpha}(\gamma_{5})_{\alpha\beta}\,\epsilon^{\beta}
+\hat{U}^{Ia\alpha}\,\bar{\epsilon}_{\beta}(\gamma_{5})^{\beta\alpha}\,\theta_{I}\nonumber\\
&&-U^{Ia\alpha}\,\hat\theta_{I}\,(\gamma_{5})^{\alpha\beta}\,\epsilon_{\beta}\,.
\end{eqnarray}

\item{The identities in the external BRST sources:}
\begin{equation}
\frac{\delta\Sigma}{\delta{\Omega^{a}_{\mu}}}=A^{a}_{\mu}\,,\qquad
\frac{\delta\Sigma}{\delta{\Lambda^{a}}}=c^{a}\,,\qquad
\frac{\delta\Sigma}{\delta{J^{a}}}=-\mathfrak{D}^{a}\,,\qquad
\frac{\delta\Sigma}{\delta{X^{a\alpha}}}=\lambda^{a}_{\alpha}\,.
\end{equation}

\item{The $U(f=4(N^2-1))$ invariance:}
\begin{eqnarray}
 \mathcal{L}^{ab}_{\mu\nu}(\Sigma) &\equiv&- \int d^{4}x \left( 
\tilde{\varphi}^{ca}_{\mu}\frac{\delta }{\delta \tilde{\varphi}^{cb}_{\nu}}
 - \varphi^{cb}_{\nu}\frac{\delta}{\delta\varphi^{ca}_{\mu}} 
+\tilde{\omega}^{ca}_{\mu}\frac{\delta }{\delta \tilde{\omega}^{cb}_{\nu}}
 - \omega^{cb}_{\nu}\frac{\delta}{\delta\omega^{ca}_{\mu}}   
+\tilde{M}^{ca}_{\sigma\mu}\frac{\delta }{\delta \tilde{M}^{cb}_{\sigma\nu}}
 - M^{cb}_{\sigma\nu}\frac{\delta}{\delta{M}^{ca}_{\sigma\mu}}  
 \right. \nonumber \\
&&\phantom{\int d^{4}x\,}  \left. 
+\tilde{N}^{ca}_{\sigma\mu}\frac{\delta }{\delta \tilde{N}^{cb}_{\sigma\nu}}
 - N^{cb}_{\sigma\nu}\frac{\delta}{\delta{N}^{ca}_{\sigma\mu}}  
\right)\Sigma = 0\;.
\end{eqnarray}
The trace of this symmetry defines  a $q_{f}$ charge and the composite index $(a,\mu)\equiv i,j,k,l,\dots$. 

\item{The $U(f'=4(N^2-1))$ invariance:}
\begin{eqnarray}
\mathcal{L}'^{ab\phantom{\alpha}\beta}_{\phantom{ab}\alpha}(\Sigma)&\equiv&\int d^{4}x\,\biggl(
\zeta^{a}_{\alpha}\frac{\delta}{\delta\zeta^{b}_{\beta}}
-\bar\zeta^{b}_{\beta}\frac{\delta}{\delta\bar\zeta^{a\alpha}}
+\theta^{a}_{\alpha}\frac{\delta}{\delta\theta^{b}_{\beta}}
-\bar\theta^{b}_{\beta}\frac{\delta}{\delta\bar\theta^{a\alpha}}
+\tilde{{V}}^{ca}_{\phantom{ca}\gamma\alpha}
\frac{\delta}{\delta\tilde{{V}}^{cb\phantom{\gamma}\beta}_{\phantom{cb}\gamma}}
-{{V}}^{cb\phantom{\gamma}\beta}_{\phantom{cb}\gamma}
\frac{\delta}{\delta{{V}}^{ca\phantom{\gamma}\alpha}_{\phantom{ca}\gamma}}
\nonumber\\
&&\phantom{\int d^{4}x\,}
+\tilde{{U}}^{ca}_{\phantom{ca}\gamma\alpha}
\frac{\delta}{\delta\tilde{{U}}^{cb\phantom{\gamma}\beta}_{\phantom{cb}\gamma}}
-{{U}}^{cb\phantom{\gamma}\beta}_{\phantom{cb}\gamma}
\frac{\delta}{\delta{{U}}^{ca\phantom{\gamma}\alpha}_{\phantom{ca}\gamma}}\biggr)\Sigma
=0\,.
\end{eqnarray}
The trace of this symmetry defines a $q_{f'}$ charge and the composite index $(a,\alpha)\equiv I,J,K,L,\dots$. 

\item{The ghost equation:}
\begin{eqnarray}
G^{a}(\Sigma)&\equiv&\int d^{4}x\,\biggl[\frac{\delta}{\delta{c}^{a}}+gf^{abc}\biggl(\check{c}^{b}\frac{\delta}{\delta{b}^{c}}
+\varphi^{b}_{i}\frac{\delta}{\delta\omega^{c}_{i}}
+\tilde\omega^{b}_{i}\frac{\delta}{\delta\tilde\varphi^{c}_{i}}
+\tilde{N}^{b}_{\mu i}\frac{\delta}{\delta\tilde{M}^{c}_{\mu i}}
+M^{b}_{\mu i}\frac{\delta}{\delta N^{c}_{\mu i}}\nonumber\\
&&\phantom{\int d^{4}x\,}+\hat{U}^{Ib\alpha}\frac{\delta}{\delta\hat{V}^{Ic\alpha}}
-U^{Ib\alpha}\frac{\delta}{\delta V^{Ic\alpha}}\biggr)\biggr]\Sigma\nonumber\\
&=&\int d^{4}x\,\left[gf^{abc}\left(K^{b}_{\mu}A^{c}_{\mu}
-L^{b}c^{c}+T^{b}\mathfrak{D}^{a}
-Y^{b\alpha}\lambda^{c}_{\alpha}\right)-\Lambda^{a}\right]\,.
\end{eqnarray}
\item{ The equation of the source $T^{a}$:}
\begin{eqnarray}
\Upsilon^{a}(\Sigma)&\equiv&\left(\frac{\delta}{\delta{T^{a}}}
+\frac{\delta}{\delta{\lambda^{a}_{\alpha}}}(\gamma_{5})_{\alpha\beta}\,\varepsilon^{\beta}
+gf^{abc} c^{b}\frac{\delta}{\delta{\mathfrak{D}^{c}}}
+gf^{abc}T^{b}\frac{\delta}{\delta{L^{c}}}\right)\Sigma\nonumber\\
&=&
3gf^{abc}\bar{\epsilon}^{\alpha}(\gamma_{\mu})_{\alpha\beta}\epsilon^{\beta}T^{b}A^{c}_{\mu} 
+ \nabla T^{b}
-gf^{abc}c^{b}J^{c}
\nonumber \\
&&
-\epsilon^{\beta}(\gamma_{5})_{\beta\alpha}X^{a\alpha}
-\bar{\epsilon}^{\alpha}(\gamma_{\mu})_{\alpha\eta}(\gamma_{5})^{\eta\beta}\epsilon_{\beta} \left( \partial_{\mu}\bar{c}^{a} + K^{a}_{\mu} \right) \;.
\end{eqnarray}
Let us also remark that some of the identities enlisted above are linearly broken, {\it i.e.} they display a breaking term which is linear in the quantum fields. Such a breaking  is a  classical breaking, not affected by the renormalization process \cite{Piguet:1995er}.

\end{itemize}

\subsubsection{Discrete symmetries}
Besides the Ward identities of the previous section, the action \eqref{fullact} is left invariant by two  useful  discrete symmetries. First, let $x_{4}\to-x_{4}$ (the same is possible for $x_{2}\to-x_{2}$). In this case we can transform the $\gamma$ matrices as
\begin{equation}
\gamma_{4}\to-\gamma_{4}\,,\qquad\gamma_{k}\to\gamma_{k},\qquad k=1,2,3\,.
\end{equation}
Notice that the anti-commutation relation $\{\gamma_{\mu},\gamma_{\nu}\}=2\delta_{\mu\nu}$ remains unchanged by the transformations above, while
\begin{equation}
\gamma_{5}\to-\gamma_{5}\,,\qquad
\mathcal{C}\to-\mathcal{C}\,,\qquad
\sigma_{4k}\to-\sigma_{4k}\,,\qquad
\sigma_{kl}\to\sigma_{kl}\,,\qquad
k,l=1,2,3\,. 
\end{equation}
Thus, given the transformations above, the action $\Sigma$ is left invariant by the following transformation of  fields and sources:
\begin{eqnarray}
\left(A^{a}_4,\mathfrak{D},T,J,M^{ab}_{4\nu},\tilde{M}^{ab}_{4\nu},N^{ab}_{4\nu},\tilde{N}^{ab}_{4\nu},K^{a}_{4},\Omega^{a}_{4}\right)
&\!\!\!\to\!\!\!& - \left(A^{a}_4,\mathfrak{D},T,J,M^{ab}_{4\nu},\tilde{M}^{ab}_{4\nu},N^{ab}_{4\nu},\tilde{N}^{ab}_{4\nu},K^{a}_{4},\Omega^{a}_{4}\right)\,,\nonumber\\
\left(\bar\lambda,\bar\epsilon,\hat\theta,\hat\zeta,Y,X\right)&\!\!\!\to\!\!\!&-i\left(\bar\lambda,\bar\epsilon,\hat\theta,\hat\zeta,Y,X\right)\,,\nonumber\\
\left(\lambda,\epsilon,\theta,\zeta\right)&\!\!\!\to\!\!\!&+i\left(\lambda,\epsilon,\theta,\zeta\right)\,.
\label{discrete1}
\end{eqnarray}
Finally, let $x_1\to-x_1$ (or $x_3\to-x_3$). In this case we have:
\begin{equation}
\gamma_{1}\to-\gamma_{1}\,,\qquad\gamma_{k}\to\gamma_{k}\,,\qquad k=2,3,4\,.
\end{equation}
Also here  the anti-commutation relation between the $\gamma$ matrices remains unchanged, while
\begin{equation}
\gamma_{5}\to-\gamma_{5}\,,\qquad
\mathcal{C}\to\mathcal{C}\,,\qquad
\sigma_{1k}\to-\sigma_{1k}\,,\qquad
\sigma_{kl}\to\sigma_{kl}\,,\qquad
k,l=2,3,4\,.
\end{equation}
Again, the action $\Sigma$ turns out to be left invariant by the following set of transformations:
\begin{equation}
\left(A^{a}_1,\mathfrak{D},T,J,M^{ab}_{1\nu},\tilde{M}^{ab}_{1\nu},N^{ab}_{1\nu},\tilde{N}^{ab}_{1\nu},K^{a}_{1},\Omega^{a}_{1}\right)
\to - \left(A^{a}_1,\mathfrak{D},T,J,M^{ab}_{1\nu},\tilde{M}^{ab}_{1\nu},N^{ab}_{1\nu},\tilde{N}^{ab}_{1\nu},K^{a}_{1},\Omega^{a}_{1}\right)\,.
\label{discrete2}
\end{equation}

\subsection{Algebraic characterization of the most general invariant counterterm}
In order to determine the most general invariant counterterm which can be freely added to each order, we follow the algebraic renormalization framework  \cite{Piguet:1995er} and perturb  the complete action $\Sigma$ by adding an integrated local polynomial in the fields and sources with dimension four and vanishing ghost number, $\Sigma_{count}$, and we require that the perturbed action, $(\Sigma + \eta \Sigma_{count})$, where $\eta$ is an infinitesimal expansion parameter, obeys the same Ward identities fulfilled by $\Sigma$ to the first order in the parameter $\eta$,. This gives the following constraints for  the counterterm $\Sigma_{count}$: 
\begin{equation}
\!\!\!\!\left.\begin{matrix}
\mathcal{B}_{\Sigma}(\Sigma_{count})=0\,,&
\displaystyle \frac{\delta}{\delta b^{a}}\,\Sigma_{count}=0\,,&
\check{G}^{a}(\Sigma_{count})=0\,,&
G^{a}(\Sigma_{count})=0\,,&
\mathcal{F}^{a}_{i}(\Sigma_{count})=0\,,&\!\!\!\!\!\!\phantom{\bigg|}\cr
\widetilde{\mathcal{F}}^{a}_{i}(\Sigma_{count})=0\,,&
\mathcal{W}^{a}_{i}(\Sigma_{count})=0\,,&
\widetilde{\mathcal{W}}^{a}_{i}(\Sigma_{count})=0\,,&
\mathcal{T}^{I}(\Sigma_{count})=0\,,&
\widehat{\mathcal{T}}^{I}(\Sigma_{count})=0\,,&\!\!\!\!\!\!\phantom{\bigg|}\cr
\displaystyle \frac{\delta}{\delta\zeta_{I}}\Sigma_{count}=0\,,&
\displaystyle\frac{\delta}{\delta\hat{\zeta}_{I}}\Sigma_{count}=0\,,&
\displaystyle\frac{\delta}{\delta\mathfrak{D}^{a}}\Sigma_{count}=0\,,&
\displaystyle \frac{\delta}{\delta\Omega^{a}_{\mu}}\Sigma_{count}=0\,&
\displaystyle\frac{\delta}{\delta\Lambda^{a}}\Sigma_{count}=0\,,&\!\!\!\!\!\!\phantom{\Bigg|}\cr
\displaystyle\frac{\delta}{\delta J^{a}}\Sigma_{count}=0\,,&
\displaystyle\frac{\delta}{\delta X^{a\alpha}}\Sigma_{count}=0\,,&
\mathcal{L}_{ij}(\Sigma_{count})=0\,,&
\mathcal{L}'^{I}_{\phantom{I}J}(\Sigma_{count})=0\,,&
\Upsilon^{a}(\Sigma_{count})=0\,.
\end{matrix}\right\}\label{constraints}
\end{equation}
where $\mathcal{B}_{\Sigma}$ is the linearized Slavnov-Taylor operator,
\begin{eqnarray}
{\cal B}_{\Sigma} &=& \int d^{4}x \biggl\{
  \biggl(\frac{\delta \Sigma}{\delta A^{a}_{\mu}}
+ \Omega^{a}_{\mu} \biggr) \frac{\delta }{\delta K^{a}_{\mu}}
+ \frac{\delta \Sigma}{\delta K^{a}_{\mu}}\frac{\delta }{\delta A^{a}_{\mu}}
+ \biggl(\frac{\delta \Sigma}{\delta \lambda^{a\alpha}}
+ X^{a\alpha}\biggr)\frac{\delta }{\delta Y^{a\alpha}}
+ \frac{\delta \Sigma}{\delta Y^{a\alpha}}\frac{\delta }{\delta \lambda^{a\alpha}}
\nonumber\\
&&
+\biggl(\frac{\delta \Sigma}{\delta c^{a}}
+ \Lambda^{a}\biggr)\frac{\delta }{\delta L^{a}} 
+ \frac{\delta \Sigma}{\delta L^{a}}\frac{\delta }{\delta c^{a}}
+ \biggl(\frac{\delta \Sigma}{\delta \mathfrak{D}^{a}}
+ {J}^{a}\biggr)\frac{\delta }{\delta T^{a}}
+ \frac{\delta \Sigma}{\delta T^{a}}\frac{\delta }{\delta \mathfrak{D}^{a}}
+ b^{a}\frac{\delta }{\delta \check{c}^{a}}
+ \omega^{a}_{i}\frac{\delta }{\delta \varphi^{a}_{i}}
+ \tilde{\varphi}^{a}_{i}\frac{\delta }{\delta \tilde{\omega}^{a}_{i}}
\nonumber\\
&&
+ \hat{\zeta}^{I}\frac{\delta }{\delta \hat{\theta}^{I}}
+ \theta^{I}\frac{\delta }{\delta \zeta^{I}}
+{V}^{Ia\alpha}\frac{\delta }{\delta {U}^{Ia\alpha}}
+ \hat{V}^{Ia\alpha}\frac{\delta }{\delta \hat{U}^{Ia\alpha}}
+ {N}^{a}_{\mu i}\frac{\delta }{\delta{M}^{a}_{\mu i}}
+ \tilde{M}^{a}_{\mu i}\frac{\delta }{\delta \tilde{N}^{a}_{\mu i}} 
+(\nabla{U}^{Ia\alpha})\frac{\delta }{\delta{V}^{Ia\alpha}}
\nonumber \\
&&
+(\nabla\hat{U}^{Ia\alpha})\frac{\delta }{\delta \hat{V}^{Ia\alpha}}
+(\nabla{M}^{a}_{\mu i})\frac{\delta }{\delta {N}^{a}_{\mu i}} 
+(\nabla\tilde{N}^{a}_{\mu i})\frac{\delta }{\delta \tilde{M}^{a}_{\mu i}}
+(\nabla K^{a}_{\mu})\frac{\delta }{\delta \Omega^{a}_{\mu}}
+(\nabla Y^{a\alpha})\frac{\delta }{\delta {X}^{a\alpha}}
\nonumber \\
&&
+(\nabla T^{a})\frac{\delta }{\delta {J}^{a}}
+(\nabla L^{a})\frac{\delta }{\delta {\Lambda}^{a}}
+(\nabla\varphi^{a}_{i})\frac{\delta }{\delta \omega^{a}_{i}}
+(\nabla\tilde{\omega}^{a}_{i})\frac{\delta }{\delta \tilde{\varphi}^{a}_{i}}
+(\nabla\hat{\theta}^{I})\frac{\delta }{\delta \hat{\zeta}^{I}}
+(\nabla\zeta^{I})\frac{\delta }{\delta \theta^{I}}  \biggr\} \;. 
\label{ST}
\end{eqnarray}
In particular, thanks to the property $\mathcal{B}_{\Sigma}\mathcal{B}_{\Sigma}=\nabla$, the general solution of the constraint $\mathcal{B}_{\Sigma}\Sigma_{count}=0$, {\it i.e.} the first eq. of \eqref{constraints},  can be written as  
\begin{equation}
\label{count}
\Sigma_{count}=a_{0}\,S_{\mathrm{SYM}}+\mathcal{B}_{\Sigma}\Delta^{(-1)}\,,
\end{equation}
where$a_0$ is a free parameter and $\Delta^{(-1)}$ is an integrated polynomial in the fields and sources of dimension 3, ghost number $-1$,  and $q_{f}=q_{f'}=0$. Taking into account the remaining constraints and the discrete symmetries \eqref{discrete1} and \eqref{discrete2}, it follows that  the most general expression for $\Delta^{(-1)}$ turns out to be 
\begin{eqnarray}
\Delta^{(-1)} &=& \int d^{4}x \bigg\{
-\frac{a_{0}}{2}\mathfrak{D}^{a}T^{a}
+ a_{1}\Big[(\partial_{\mu}\check{c}^{a}+K^{a}_{\mu})A^{a}_{\mu}
+ M^{a}_{\mu i}D^{ab}_{\mu}\tilde{\omega}^{b}_{i}
+ (\partial_{\mu}\tilde{\omega}^{a}_{i})D^{ab}_{\mu}\varphi^{b}_{i}
\nonumber\\
&&
+ \tilde{N}^{a}_{\mu i}M^{a}_{\mu i}
+ \tilde{N}^{a}_{\mu i}D^{ab}_{\mu}\varphi^{b}_{i}\Big]
+ a_{2}\left(Y^{a\alpha}-\hat{U}^{Ia}_{\phantom{a}\beta}\mathcal{C}^{\alpha\beta}\theta_{I}-U^{Ia\alpha}\hat{\theta}_{I}\right)\lambda^{a}_{\alpha}
\nonumber\\
&&
+\left(\frac{a_{0}}{2} -a_{2}\right)\left(Y^{a\alpha}-\hat{U}^{Ia}_{\phantom{a}\beta}\mathcal{C}^{\alpha\beta}\theta_{I}-U^{Ia\alpha}\hat{\theta}_{I}\right)(\gamma_{5})_{\alpha\gamma}\epsilon^{\gamma}T^{a}
\bigg\}\;.
\end{eqnarray}
with $a_1,a_2$ free parameters. We see therefore that  $\Sigma_{count}$ depends on three arbitrary coefficients, {\it i.e.}  $(a_0, a_1, a_{2})$. 
\subsection{Renormalization factors}
In order to complete the analysis of the algebraic renormalization, we still need to show that the counterterm $\Sigma_{count}$ can be reabsorbed into the starting action $\Sigma$ through a redefinition of the fields $\{\phi \}$,  sources $\{ S \}$ and   parameters $\{p\}=\{g,\epsilon\}$,  namely, 
\begin{equation}
\label{ration}
\Sigma(\phi,S,g) + \eta \Sigma_{count}(\phi,S,p)  = \Sigma(\phi_0,S_0,p_0) + \mathcal{O}(\eta^2) \;, 
\end{equation}
where $(\phi_0, S_0, p_0)$ are the so-called bare quantities, defined through the renormalization factors as 
\begin{equation}
\label{renormfs}
\phi_{0}=Z^{1/2}_{\phi}\,\phi\,,   \qquad
S_{0}=Z_{S}\,S\,,  \qquad p_0 = Z_p p\,, 
\end{equation}
where
\begin{equation}
Z^{1/2}_{\phi}=1+\eta \frac{z_{\phi}}{2}+\mathcal{O}(\eta^{2})\,,\qquad
Z_{S}=1+\eta\,z_S+\mathcal{O}(\eta^{2})\,, \qquad Z_p = 1 +\eta z_p+\mathcal{O}(\eta^{2})  \;, 
\end{equation}
with the $\{z \}$ being  linear combinations of the coefficients $(a_0,a_1,a_2)$. Moreover, in the present case, a little care has to be taken due to the  potential mixing of quantities which have the same quantum numbers, see also \cite{Capri:2014xea,Capri:2014jqa}.  In fact,  as it can be checked from Table \ref{tabfields} and Table \ref{tabsources}, one  notices  that the field $\lambda^{a\alpha}$ and the combination $\gamma_{5}\epsilon T^{a}$ have the same dimension and quantum numbers as well as the field $\mathfrak{D}^{a}$ and the combination $(Y^{a}-\hat{U}^{Ia}\mathcal{C}\theta_{I}-U^{Ia}\hat{\theta}_{I})\gamma_{5}\epsilon$.  As a consequence, these quantities can mix at quantum level, a well known property of renormalization theory. This feature can be properly taken into account by writing the renormalization of the fields $\lambda$ and $\mathfrak{D}$  as  
\begin{equation}
\label{lrenorm}
\lambda^{a\alpha}_{0}=Z^{1/2}_{\lambda}\,\lambda^{a\alpha}+\eta\, z_{1}\,T^{a}(\gamma_{5})^{\alpha\beta}\epsilon_{\beta}+\mathcal{O}(\eta^{2})
\end{equation}
and
\begin{equation}
\label{drenorm}
\mathfrak{D}^{a}_{0}=Z^{1/2}_{\mathfrak{D}}\,\mathfrak{D}^{a}+\eta\,z_2\left(Y^{a\alpha}-\hat{U}^{Ia}_{\phantom{a}\beta}\mathcal{C}^{\alpha\beta}\theta_{I}-U^{Ia\alpha}\hat{\theta}_{I}\right)(\gamma_{5})_{\alpha\gamma}\epsilon^{\gamma}+\mathcal{O}(\eta^{2})\,,
\end{equation}
while the remaining fields, sources and parameters still obey \eqref{renormfs}.\\\\  By direct inspection of eq.\eqref{ration}, we find 
\begin{eqnarray}
Z^{1/2}_{A} &=& 1 + \eta\left(\frac{a_{0}}{2}+a_{1}\right)+\mathcal{O}(\eta^{2})\;, \nonumber \\
Z^{1/2}_{\lambda}& = &1+ \eta \left(\frac{a_{0}}{2}-a_{2}\right)+\mathcal{O}(\eta^{2})\;, \nonumber \\
Z_{g}&=&1-\eta\frac{a_{0}}{2}+\mathcal{O}(\eta^{2})\;.
\end{eqnarray}
All other remaining renormalization factors can be expressed in terms of the tree independent quantities $(Z_A, Z_\lambda, Z_g)$, namely 
\begin{eqnarray}
&&
Z_{\mathfrak{D}}^{1/2}=1\;, \nonumber \\
&&
Z^{1/2}_{\bar{\varphi}} = Z^{1/2}_{\varphi} = Z^{1/2}_{c}=Z^{1/2}_{\check{c}} = Z_{K} = Z^{-1/2}_{g}Z^{-1/4}_{A}\;, \nonumber \\
&&
Z^{1/2}_{\bar{\omega}} = Z^{-1}_{g}\;, \nonumber \\
&&
Z^{1/2}_{\omega} = Z^{-1/2}_{A} \nonumber \\
&&
Z^{1/2}_{\theta}=Z^{1/2}_{\hat{\theta}} = 1\;, \nonumber \\
&&
Z^{1/2}_{\zeta}=Z^{-1/2}_{\hat{\zeta}} = Z^{1/2}_{g}Z^{-1/4}_{A}\,.
\end{eqnarray}
In particular, from eq.\eqref{physval1}, it follows that the renormalization of the sources $M$ and $\tilde{M}$ gives us the renormalization factor of the Gribov parameter $\gamma^{2}$, while the renormalization of $V$ and $\hat{V}$ yields the renormalization of $M^{3/2}$, {\it i.e.}
\begin{eqnarray}
&&Z_{\gamma^{2}}=Z_{\tilde{M}} =Z_{M} = Z^{-1/2}_{g}Z^{-1/4}_{A}\;, \nonumber \\
&&Z_{M^{3/2}}=Z_{\hat{V}} =Z_{V} = Z^{-1/2}_{\lambda}\;.
\end{eqnarray}
The other sources renormalize as
\begin{eqnarray}
&&Z_{N} = Z_\Omega=Z^{-1/2}_{A}\;, \nonumber \\
&&Z_{\bar{N}} = Z^{-1}_{g}\;, \nonumber \\
&&Z_U=Z_{\hat{U}}=Z_Y = Z^{-1/2}_{g}Z^{1/4}_{A}Z^{-1/2}_{\lambda}\;, \nonumber \\
&&Z_{L} = Z^{1/2}_{A}\;, \nonumber \\
&&Z_{T} = Z_{\Lambda}=Z^{1/2}_{g}Z^{1/4}_{A}\;, \nonumber \\
&&Z_{X} = Z^{-1/2}_{\lambda}\;, \nonumber \\
&&Z_{J} = 1\,.
\end{eqnarray}
Finally, the renormalization factor of the supersymmetric ghost  parameter $\epsilon$ is
\begin{equation}
Z_{\epsilon}=Z^{1/2}_{g}Z^{-1/4}_{A} \;, 
\end{equation}
while we also have 
\begin{equation}
z_{1} =- z_{2} = -\frac{a_{0}}{2} + a_{2}\;. 
\end{equation}
This concludes the proof of the algebraic renormalizability of the model.
\section{Notations and conventions in Euclidean space-time}
\label{notations}
\noindent
\textbf{Units}: $\hbar=c=1$.

\noindent
\textbf{Euclidean metric}: $\delta_{\mu\nu}=diag(+,+,+,+)$.

\noindent
\textbf{Wick rotations:} $x_0\rightarrow -ix_4\Rightarrow\partial_0\rightarrow+i\partial_4$, $A_0\rightarrow+iA_4$.

\noindent
\textbf{Gamma matrices:}
$$
\gamma_4=\left( \begin{array}{cc}
0 & \mathbf{1} \\
\mathbf{1} & 0 \end{array} \right)\,,\qquad
\gamma_k=-i\left( \begin{array}{cc} 0 & \sigma_k \\ -\sigma_k & 0 \end{array} \right)\,.
$$

\noindent
\textbf{Pauli matrices}:
$$
\sigma_4=\left( \begin{array}{cc}
1 & 0 \\
0 & 1 \end{array} \right),\qquad
\sigma_1=\left( \begin{array}{cc} 0 & 1\\ 1 & 0 \end{array} \right)\,,\qquad
\sigma_2=\left( \begin{array}{cc} 0 & -i\\ i & 0 \end{array} \right)\,,\qquad
\sigma_3=\left( \begin{array}{cc} 1 & 0 \\ 0 & -1 \end{array} \right)\,.
$$

\noindent
The  gamma matrices obey the following  relations:
\begin{eqnarray}
 \gamma_\mu&=&\gamma_\mu^\dagger\,,\\
  \{\gamma_\mu,\gamma_\nu\}&=&2\delta_{\mu\nu}\,.
\end{eqnarray}

\noindent
We also define the  matrix $\gamma_5$ as
$$
\gamma_5=\gamma_4\gamma_1\gamma_2\gamma_3=\left(\begin{array}{cc}
                                                   \mathbf{1} & 0\\
                                                   0 & -\mathbf{1}\\
                                                  \end{array}\right)\,,
$$

\noindent
with the following properties:
\begin{equation}
 \{\gamma_5,\gamma_\mu\}=0\,,\qquad(\gamma_5)^2=\mathbf{1}\,,\qquad\gamma_5^\dagger=\gamma_5\,.
\end{equation}

\noindent
The charge conjugation matrix is
\begin{equation}
\label{Cmtrx}
\mathcal{C}=\gamma_4\gamma_2=i\left(\begin{array}{cc} \sigma_2 & 0 \\ 0 & -\sigma_2\\ \end{array}\right)\,,
\end{equation}

\noindent
with 
\begin{equation}
 \mathcal{C}^{-1}=-\mathcal{C}=\mathcal{C}^T\,,\qquad\mathcal{C}^{-1}\gamma_\mu\mathcal{C}=-
\gamma_\mu^T\,.
\end{equation}

\noindent
The matrices $\sigma^{\mu\nu}$ are  defined as
\begin{equation}
  (\sigma_{\mu\nu})^{~\beta}_\alpha\equiv\frac{1}{2}[\gamma_\mu,\gamma_\nu]^{~\beta}_\alpha
\end{equation}
and  have the property $\sigma_{\mu\nu}^\dagger=-\sigma_{\mu\nu}$.

\noindent
\textbf{Majorana fermions:}

\noindent
The Majorana condition reads:
\begin{equation}
\label{mjconj}
 \lambda^\mathcal{C}=\lambda=\mathcal{C}\bar{\lambda}^T\qquad
\Longleftrightarrow\qquad\bar{\lambda}=\lambda^T\mathcal{C}\;,
\end{equation}
leading to the following relations
\begin{equation}
\bar{\lambda}\gamma_{\mu}\epsilon = \bar{\epsilon}\gamma_{\mu}\lambda \qquad \text{and} \qquad \bar{\lambda}\gamma_{\mu}\gamma_{5}\epsilon = - \bar{\epsilon}\gamma_{\mu}\gamma_{5}\lambda\;.
\end{equation}

\noindent
\textbf{Fierz identity in Euclidean space-time:}
\begin{eqnarray}
\epsilon_{1}\bar{\epsilon}_{2} &=&  \frac{1}{4}(\bar{\epsilon}_{2}\epsilon_{1})\mathbf{1} 
+ \frac{1}{4}(\bar{\epsilon}_{2}\gamma_{5}\epsilon_{1})\gamma_{5}
+ \frac{1}{4}(\bar{\epsilon}_{2}\gamma_{\mu}\epsilon_{1})\gamma_{\mu}
- \frac{1}{4}(\bar{\epsilon}_{2}\gamma_{\mu}\gamma_{5}\epsilon_{1})\gamma_{\mu}\gamma_{5}\nonumber \\
&&
- \frac{1}{8}(\bar{\epsilon}_{2}\sigma_{\mu\nu}\epsilon_{1})\sigma_{\mu\nu}    \;.
\end{eqnarray}


\noindent {\bf Indices notations}:
We display here a summary of the indices used in Appendix \ref{algrenorm} 
\begin{center}
\begin{tabular}{ll}
$\bullet$&The Lorentz indices: $\mu,\nu,\rho,\sigma,\lambda\in\{1,2,3,4\}$\,;$\phantom{\Bigl|}$\\
$\bullet$&The Spinor indices: $\alpha,\beta,\gamma,\delta,\eta\in\{1,2,3,4\}$\,;$\phantom{\Bigl|}$\\
$\bullet$&The $SU(N)$ group indices: $a,b,c,d,e\in\{1,\dots,N^{2}-1\}$\,;$\phantom{\Bigl|}$\\
$\bullet$&The composite-index $(a,\mu)$: $i,j,k,l\in\{1,\dots,f=4(N^{2}-1)\}$\,;$\phantom{\Bigl|}$\\
$\bullet$&The composite-index $(a,\alpha)$: $I,J,K,L\in\{1,\dots,f'=4(N^{2}-1)\}$\,.$\phantom{\Bigl|}$\\
\end{tabular}
\end{center}

\noindent {\bf Table of quantum numbers}:
We display below the quantum numbers of the fields and sources appearing in the action \eqref{fullact}. Notice that by ``nature'' we mean ``C" for commuting (or bosonic) and ``A"  for anti-commuting (or fermionic). 
\begin{table}[h!]
\begin{tabular}{l|c|c|c|c|c|c|c|c|c|c|c|c|c|c|c|c}
\hline
\phantom{\Big|}
&$A$&$\lambda$&$\mathfrak{D}$&$c$&$\check{c}$&$b$&$\varphi$&$\tilde\varphi$&$\omega$&$\tilde\omega$&$\zeta$&$\hat{\zeta}$&$\theta$&$\hat{\theta}$
&$\epsilon$&$\bar{\epsilon}$\cr
\hline
$\phantom{\Bigl|}\!\!$Dimension
&1&$\frac{3}{2}$&2&1&1&2&1&1&2&0&0&2&1&1&$\frac{1}{2}$&$\frac{1}{2}$\cr
\hline
$\phantom{\Bigl|}\!\!$Ghost\#
&0&0&0&1&$-1$&0&0&0&1&$-1$&$-1$&1&0&0&1&1\cr
\hline
$\phantom{\Bigl|}\!\!$Charge-$q_f$
&0&0&0&0&0&0&1&$-1$&1&$-1$&0&0&0&0&0&0\cr
\hline
$\phantom{\Bigl|}\!\!$Charge-$q_{f'}$
&0&0&0&0&0&0&0&$0$&0&$0$&1&$-1$&1&$-1$&0&0\cr
\hline
$\phantom{\Bigl|}\!\!$Nature
&C&A&C&A&A&C&C&C&A&A&C&C&A&A&C&C\cr
\hline
\end{tabular}
\caption{The quantum numbers of fields.}
\label{tabfields}
\end{table}

\begin{table}[h!]
\begin{tabular}{l|c|c|c|c|c|c|c|c|c|c|c|c|c|c|c|c}
\hline
\phantom{\Big|}&$U$&$\hat{U}$&$V$&$\hat{V}$&$M$&$\tilde{M}$&$N$&$\tilde{N}$&$K$&$\Omega$&$L$&$\Lambda$&$T$&$J$&$Y$&$X$\cr
\hline
$\phantom{\Bigl|}\!\!$Dimension
&$\frac{1}{2}$&$\frac{1}{2}$&$\frac{3}{2}$&$\frac{3}{2}$&2&2&3&1&2&3&2&3&1&2&$\frac{3}{2}$&$\frac{5}{2}$\cr
\hline
$\phantom{\Bigl|}\!\!$Ghost\#
&$-1$&$-1$&0&0&$0$&0&1&$-1$&$-1$&$0$&$-2$&$-1$&$-1$&0&$-1$&0\cr
\hline
$\phantom{\Bigl|}\!\!$Charge-$q_f$
&0&0&0&0&1&$-1$&1&$-1$&0&$0$&0&0&0&0&0&0\cr
\hline
$\phantom{\Bigl|}\!\!$Charge-$q_{f'}$
&1&$-1$&1&$-1$&0&0&0&$0$&0&$0$&0&$0$&0&$0$&0&0\cr
\hline
$\phantom{\Bigl|}\!\!$Nature
&A&A&C&C&C&C&A&A&A&C&C&A&A&C&C&A\cr
\hline
\end{tabular}
\caption{The quantum numbers of external sources.}
\label{tabsources}
\end{table}

\end{appendix}

----------------------------------------------------------------------------------------
\section*{Acknowledgments}

The Conselho Nacional de Desenvolvimento Cient\'{\i}fico e
Tecnol\'{o}gico (CNPq-Brazil), the Faperj, Funda{\c{c}}{\~{a}}o de
Amparo {\`{a}} Pesquisa do Estado do Rio de Janeiro, the SR2-UERJ,  the
Coordena{\c{c}}{\~{a}}o de Aperfei{\c{c}}oamento de Pessoal de
N{\'{\i}}vel Superior (CAPES)  are gratefully acknowledged.

\end{document}